\pdfoutput=1

\documentclass[a4paper,fleqn,usenatbib]{article}

\usepackage{amsmath}	% Advanced maths commands
\usepackage{abbreviations}

\usepackage{amssymb}	% Extra maths symbols
\usepackage{txfonts}
\usepackage{natbib}
\usepackage{hyperref}
\usepackage[mathcal]{euscript}

\usepackage[T1]{fontenc}
\usepackage{ae,aecompl}

\usepackage[a4paper,margin=2.5cm]{geometry}
\usepackage{graphicx}	% Including figure files
\usepackage{subfig}
\usepackage{color}

\usepackage[utf8]{inputenc}
\usepackage{authblk}

\DeclareMathAlphabet{\mathpzc}{OT1}{pzc}{m}{it}

\begin{document}
\title{Generalized, energy-conserving numerical simulations of particles in general relativity. I. Time-like and null geodesics}
\author[1]{F. Bacchini\thanks{E-mail: fabio.bacchini@kuleuven.be}}
\author[1]{B. Ripperda}
\author[2]{A.Y. Chen}
\author[3]{L. Sironi}
\affil[1]{Centre for mathematical Plasma Astrophysics, Department of Mathematics, KU Leuven, Celestijnenlaan 200B, B-3001 Leuven, Belgium}
\affil[2]{Department of Astrophysical Sciences, Princeton University, Peyton Hall, Ivy Lane, Princeton, NJ 08544, USA}
\affil[3]{Department of Astronomy, Columbia University, 550 W 120th St, New York, NY 10027, USA}
\renewcommand\Authands{ and }

\label{firstpage}
\maketitle

\begin{abstract}
The numerical integration of particle trajectories in curved spacetimes is fundamental for obtaining realistic models of the particle dynamics around massive compact objects such as black holes and neutron stars. Generalized algorithms capable of handling generic metrics are required for studies of both standard spacetimes (Schwarzschild and Kerr metrics) and non-standard spacetimes (e.g. Schwarzschild metric plus non-classical perturbations or multiple black hole metrics). The most commonly employed explicit numerical schemes (e.g. Runge-Kutta) are incapable of producing highly accurate results at critical points, e.g. in the regions close to the event horizon where gravity causes extreme curvature of the spacetime, at an acceptable computational cost. Here, we describe a generalized algorithm for the numerical integration of time-like (massive particles) and null (photons) geodesics in any given 3+1 split spacetime. We introduce a new, exactly energy-conserving implicit integration scheme based on the preservation of the underlying Hamiltonian, and we compare its properties with a standard fourth-order Runge-Kutta explicit scheme and an implicit midpoint scheme. We test the numerical performance of the three schemes against analytic solutions of particle and photon orbits in Schwarzschild and Kerr spacetimes. We also prove the versatility of our framework in handling more exotic metrics such as Morris-Thorne wormholes and quantum-perturbed Schwarzschild black holes. The generalized approach is also discussed in the perspective of future extensions to more complex particle dynamics, e.g. the addition of the Lorentz force acting on charged particles, which allows for test particle diagnostics in GRMHD simulations.
\end{abstract}


%%%%%%%%%%%%%%%%%%%%%%%%%%%%%%%%%%%%%%%%%%%%%%%%%%

%%%%%%%%%%%%%%%%% BODY OF PAPER %%%%%%%%%%%%%%%%%%

\section{Introduction}

\label{sec:intro}
To model the physics of accreting black holes and neutron stars, an accurate description of particle motion in general relativity (GR) is essential. In absence of electromagnetic forces, particles follow geodesic paths in curved spacetimes. Geodesic integrators are gaining much attention in the context of ray-tracing simulations (\citealt{kuchelmeister2012}; \citealt{pu2016}; \citealt{vincent2016}; \citealt{chan2017}), to describe radiation effects in general relativistic magnetohydrodynamic (GRMHD) simulations (e.g. \citealt{mckinney2014}; \citealt{white2016}; \citealt{narayan2017}; \citealt{porth2017}), as well as to compare to observational results (\citealt{doeleman2012}; \citealt{boehle2016}; \citealt{Pu2017}). To this end, several approaches and tools are available to the community. Elaborate graphical interfaces even allow for interactive 3D representations of geodesic bundles in curved spacetimes (\citealt{mullergrave2010}), while optimized parallel algorithms result in extremely fast simulations on modern architectures (\citealt{chan2017}). However, modern observational developments in the study of the very nature of compact objects, such as black holes, call for further exploration, and possibly refinement, of the available numerical methods. In this paper, we present an optimized formulation of a generalized algorithm addressing i) the need to consider generic spacetimes, including exotic spacetimes differing from the classical Schwarzschild and Kerr metrics, ii) the possibility of extending the same algorithm to include external (conservative) forces, e.g. the Lorentz force on charged particles, to trace particles in electromagnetic (EM) fields (e.g. from GRMHD simulations), and iii) the need to retain the highest possible accuracy in the results with a reasonable computational cost.

The geodesic equations of motion are a set of 4 (in 4-dimensional spacetime) second-order, nonlinear differential equations. Due to their nature, the form of the equations depends on the underlying spacetime metric. Many codes carry out calculations specialized for one single metric, or utilize analytic solutions where these are available (e.g. for Schwarzschild and Kerr spacetimes). Although this approach certainly improves the resulting performance, it also limits the freedom to apply a given algorithm to different physical regimes (e.g. adding spin to a Schwarzschild metric), handle perturbations of standard spacetimes, or deal with numerically defined metrics, or in general with non-integrable spacetimes. Recent works on ray-tracing and GRMHD calculations indeed considered non-standard spacetimes (\citealt{younsi2016}; \citealt{mizuno2018}), in view of comparing upcoming observations with simulation results. In this case, a general approach to the solution of the geodesic equation is required, since it allows for simulations in different spacetimes without altering the code structure, and it deals with those metrics where analytic solutions are not available. When computational schemes are used to solve the geodesic equations of motion, a common choice is the application of explicit numerical methods, such as high-order Runge-Kutta integrators (\citealt{mullergrave2010} \citealt{vincent2011}; \citealt{baubock2012}; \citealt{psaltis2012}; \citealt{chan2017}). These methods are generally very accurate for evolving neutral massive particles and photons. They are, however, characterized by intrinsic non-conservation of the invariants of motion, which results in a secular, unphysical energy drift, that typically compromises the long-term stability of the overall scheme. Moreover, in the plasma surrounding compact objects, external forces acting on charged particles e.g. due to electromagnetic fields have to be taken into consideration. Non-symplectic/geometric methods e.g. explicit Runge-Kutta are then generally discarded due to their intrinsic unbounded accumulation of energy errors, as well as the incapability of accurately grasping periodic motion in such fields over long times (\citealt{Qin_2013}).

In this work, we present a proof-of-concept implementation of a new geodesics integrator. In our formulation we have the freedom of resolving the motion of free particles (i.e. particles traveling on geodesics) in any given spacetime, provided that the 3+1 split metric is available as input. We include the option of using a Runge-Kutta method of desired order, as well as a second-order implicit midpoint scheme. Additionally, we introduce a new second-order implicit scheme, which is an extension of the energy-conserving special relativistic implicit integrator described in \cite{LapentaMarkidis2011} and \cite{ripperda2018}. The new scheme is based on the discretization of the underlying Hamiltonian that describes the particle motion in curved spacetime. This Hamiltonian method is energy-conserving (to round-off precision) by design. We show that this property allows for treating spacetime regions that are pathological to the standard Runge-Kutta and implicit midpoint methods, without extreme reduction of the integration step. Long-time stability of the simulated trajectories is achieved, allowing for the calculation of photon and particle orbits over arbitrarily long periods. The resulting numerical scheme retains a highly general character, such that future inclusions of external forces (e.g. the Lorentz force for charged particles) remain straightforward, and will be treated in follow-up works.

All methods are implemented such that they can handle any metric spacetime in four dimensions. We test the three integrators in standard spacetimes describing black holes, like the Schwarzschild metric for static black holes and the Kerr metric for spinning black holes. As an example of the versatility of the framework presented here, we also integrate geodesics in a Morris-Thorne wormhole, in a Reissner-Nordstr\"{o}m dihole, and in a Schwarzschild metric that is perturbed by microscopic effects. This provides the opportunity to compare trajectories in such exotic spacetimes and potentially obtain observables for upcoming results from the Event Horizon Telescope. Our results are comparable to previous works that presented frameworks for the integration of geodesics (\citealt{chan2017}; \citealt{takahashiumemura2017}; \citealt{bronzwaer2018}; \citealt{moscibrodzkagammie2018}). However, here we introduce the flexibility of extending our code to include external (conservative) forces, and we focus on advanced numerical methods that attain energy conservation.

In Section \ref{sec:theory}, we briefly review the set of equations and the 3+1 split formulation adopted in this work. In Section \ref{sec:schemes} we describe all numerical methods and their characteristics. In Section \ref{sec:tests} we test the accuracy of all methods in the two most commonly used spacetimes: the Schwarzschild metric, describing static black holes, and the Kerr metric, describing spinning black holes. We evolve photons and massive particles on analytically known geodesic orbits and compare errors on the position and the energy of the particle. As a proof-of-principle for future developments, we also show applications in more exotic spacetimes that are candidates to describe quantum effects around black holes in Section \ref{sec:applications}. In Section \ref{sec:discussionsummary} we summarize our results and conclusions and we give an outlook on future work.

\section{General relativistic test particles}
\label{sec:theory}
The motion of test particles in general relativity is described by the geodesic equation (see e.g. \citealt{carrol})
\begin{equation}
\frac{d^2x^{\mu}}{d\tau^2} + \Gamma^{\mu}_{\lambda\sigma} \frac{d x^{\lambda}}{d \tau} \frac{dx^{\sigma}}{d\tau} = 0,
\label{eq:geodesic}
\end{equation}
where $\mu=0,1,2,3$ and $\Gamma^\mu_{\lambda\sigma}$ is the Christoffel connection for a Riemannian metric. The derivative of the four-position $x^\mu$ with respect to an affine parameter $\tau$ is the contravariant four-velocity,
\begin{equation}
 \frac{dx^\mu}{d\tau}=u^\mu,
\end{equation}
where we choose, for the remainder of this work, $t:=x^0$ and hence $u^0=dt/d\tau$ in units where $c=1$. The spatial components of the four-position are indicated as $x^i$, with $i=1,2,3$.

In the context of numerical integration, it is common practice to rewrite the equations above in the framework of the Arnowitt-Deser-Misner (ADM) formalism (e.g. \citealt{rezzollazanotti}). In this formulation, any metric can be written in the form
\begin{equation}
 g_{\mu\nu}=
 \begin{pmatrix}
  -\alpha^2+\beta_k\beta^k & \beta_i \\
  \beta_j & \gamma_{ij}
 \end{pmatrix},
 \label{eq:3p1metric}
\end{equation}
where $\alpha$ is the lapse function, $\beta^i$ is the shift three-vector, and $\gamma_{ij}$ is the spatial part of $g_{\mu\nu}$. The procedure through which the so-called 3+1 split metric in equation (\ref{eq:3p1metric}) is obtained results in a foliation of spacetime into space-like hypersurfaces of constant coordinate time $t$. The corresponding inverse metric reads
\begin{equation}
 g^{\mu\nu}=
 \begin{pmatrix}
  -1/\alpha^2 &\beta^i/\alpha^2 \\
  \beta^j/\alpha^2 & \gamma^{ij}-\beta^i\beta^j/\alpha^2
 \end{pmatrix},
 \label{eq:3p1invmetric}
\end{equation}
where $\gamma^{ij}$ is the algebraic inverse of $\gamma_{ij}$, and $\beta^i=\gamma^{ij}\beta_j$. In this work, we choose a $(-,+,+,+)$ signature for the metric in equation (\ref{eq:3p1metric}). Finally, the corresponding line element is written,
\begin{equation}
 ds^2=-\alpha^2 dt^2 + \gamma_{ij}(dx^i+\beta^i dt)(dx^j+\beta^j dt),
\end{equation}
hence it is generally straightforward to obtain the expressions of $\alpha$, $\beta^i$ and $\gamma_{ij}$ from the standard formulation of any common metric, e.g. the Schwarzschild or Kerr metrics.

With the definitions above, the geodesic equation (\ref{eq:geodesic}) can be rewritten in terms of first-order evolution equations in the variables $x^i$ and $u_i=g_{i\mu}u^\mu$, such that
\begin{equation}
\frac{d x^i}{dt} = \gamma^{ij} \frac{u_j}{u^0} - \beta^i,
\label{eq:geodesic3p1x}
\end{equation}
\begin{equation}
\frac{du_i}{dt} = -\alpha u^0 \partial_i \alpha + u_k \partial_i \beta^k - \frac{u_j u_k}{2 u^0} \partial_i \gamma^{jk},
\label{eq:geodesic3p1u}
\end{equation}
where 
\begin{equation}
u^0 = \left(\gamma^{jk} u_j u_k+\epsilon\right)^{1/2}/\alpha,
\label{eq:lfac}
\end{equation}
with $\epsilon=0$ for null geodesics (i.e. photon paths) and $\epsilon=1$ for time-like geodesics (i.e. free-falling particle orbits).

The system of equations (\ref{eq:geodesic3p1x}-\ref{eq:geodesic3p1u}) is suitable for numerical integration in a number of ways. First, as noted above, the equations are first-order, hence reducing the complexity of the required numerical scheme. Second, there is no need to integrate temporal components, reducing the number of equations from 8 to 6, thanks to the definition of $u^0$ from equation (\ref{eq:lfac}). Furthermore, it can be shown that such definition of $u^0$ enforces the conservation of the norm of the four-velocity, $u^\mu u_\mu=-\epsilon$. Finally, integrating in coordinate time $t$ rather than in proper time $\tau$ makes it easier to embed the motion of test particles in the time evolution of global fields, e.g. as obtained from GRMHD codes such as {\tt BHAC} (\citealt{porth2017}). Aside from the numerical methods presented in the next sections, in this work we aim at testing also the versatility of this formulation applied to different spacetimes, including nonstandard ones. In this context, we purposely generalize the implementation of the numerical schemes such that, given the spacetime functions $\alpha$, $\beta^i$, and $\gamma^{ij}$, the basic steps of the algorithm remain fixed.

In this work, we consider stationary metrics with no dependence on coordinate time $t$, hence $\alpha$, $\beta^i$, and $\gamma^{ij}$ are functions of $x^i$ only. As a direct consequence, in all cases there exists at least one Killing vector $K=\partial_t$ representing the symmetry of the metric with respect to translations in time. This defines a conserved quantity $-K^\mu u_\mu=E$ which we label as the energy of the test particle. By definition, since $K^\mu=(1,0,0,0)$, we have the correspondence $E=-u_0$. Conservation of energy characterizes all metrics discussed in the next sections, and is an important physical aspect that has to be ensured as closely as possible in numerical integration of particle and photon trajectories. Other conserved quantities may or may not exist, depending on the metric considered.

\section{Numerical schemes}
\label{sec:schemes}
In this section, we recap the main properties of the numerical integrators considered in this work. We also introduce an energy-conserving, second-order implicit scheme derived from the Hamiltonian of the system of equations (\ref{eq:geodesic3p1x}-\ref{eq:geodesic3p1u}). A brief discussion on the computational cost of the three schemes can be found in Appendix \ref{app:compcost}.

In this work we focus on simulating free particles traveling on geodesics, thus neglecting external forces. However, it is important to keep the algorithm as general as possible, in the perspective of extending to more complicated physics. In particular, in the context of particle-based simulations of kinetic dynamics (\citealt{spitkovsky}) or test particles in relativistic macroscopic flows (\citealt{ripperda}), electromagnetic forces must be included. Hence, in our analysis of integration schemes, we must take into account the behavior of each algorithm in the context of the aforementioned fields of application, in addition to pure geodesic motion.

\subsection{Explicit integration schemes}
\label{sec:rk4}

Explicit methods for ordinary differential equations (ODEs) come in many different forms and present a number of advantageous properties, the most remarkable being their capability of advancing the numerical solution in a finite number of non-iterative steps. The most widely used explicit method in scientific computing is the fourth-order Runge-Kutta method (RK4 from now on) (see e.g. \citealt{press}). For a discretization in time $t$ of an ODE problem such as
\begin{equation}
 \frac{y^{n+1}-y^n}{\Delta t}=f(t,y(t)),
 \label{eq:discode}
\end{equation}
the RK4 scheme advances the variable $y(t)$ by means of the steps
\begin{equation}
\begin{aligned}
 k_1 & = \Delta t f(t^n,y^n), \\
 k_2 & = \Delta t f(t^n+\Delta t/2,y^n+k_1/2), \\
 k_3 & = \Delta t f(t^n+\Delta t/2,y^n+k_2/2), \\
 k_4 & = \Delta t f(t^n+\Delta t,y^n+k_3), \\
 y^{n+1} & =y^n+\frac{1}{6}(k_1+2k_2+2k_3+k_4).
\end{aligned}
\end{equation}
At each time step, one needs to evaluate the right-hand side of equation (\ref{eq:discode}) four times. The resulting error in $y(t)$ in the computed solution is of order $O(\Delta t^5)$.

All explicit methods introduce, aside from errors of a given order in the solution variable $y(t)$, errors in other properties of the real dynamical system described by the associated differential equation. In particular, explicit methods are generally (but not in all cases) incapable of preserving first integrals of the system, such as the associated Hamiltonian, if this exists. Another well-known issue of explicit schemes is the non-conservation of phase space volume (see e.g. \citealt{hairer}; \citealt{fengqin}). The error in these quantities decreases if the time step size is reduced, but it inevitably accumulates over time. For some cases, especially over long times, the resulting computed solution becomes unacceptably inaccurate. However, for a widely used method as the RK4 scheme introduced above, the scaling of errors with the reduction of $\Delta t$ is satisfactory enough to be generally acceptable. Refinements of standard explicit methods, such as adaptive step control (e.g. \citealt{gear}) are also widely employed, although the overall computational cost is increased. Furthermore, whenever an explicit method of a given order proves unsatisfactory, one has the freedom to increase the order of the method at the cost of complicating the solution procedure.

\subsection{Symplectic implicit schemes}
Whenever standard explicit schemes are not suitable for a specific problem, more advanced, often implicit integration schemes prove necessary. Here, we consider addressing the issue of preserving trajectories in phase space, a feature that simple explicit schemes such as the RK4 scheme) often lack. The conservation of phase space properties (trajectories and volume) characterizes symplectic schemes. A symplectic integration scheme is such that it conserves a ``modified Hamiltonian'' (\citealt{springel2014}), that differs from the real Hamiltonian of the considered differential equation. Because this holds regardless of the time step, the long-term behavior of such integrators is generally superior to non-symplectic explicit schemes. Advanced explicit symplectic schemes (\citealt{liu2016}), as well as implicit schemes with adaptive step size control (\citealt{seyrichlukesgerakopoulos2012}) have been successfully applied to Hamiltonian systems describing particle motion in curved spacetimes. Despite of the order of accuracy, however, symplectic schemes present essentially the same characteristics, with long-term conservation of the first integrals of the motion to a level of accuracy that scales with the time step and the order of the method. Hence, for simplicity, here we restrict our comparisons to the implicit midpoint rule (IMR from now on) discretization scheme, which is the simplest second-order implicit symplectic integrator (\citealt{feng1986}). Applied to the example ODE (\ref{eq:discode}), the scheme advances the variable $y(t)$ according to
\begin{equation}
 y^{n+1}=y^n+\Delta t f\left(\frac{t^n+t^{n+1}}{2},\frac{y^n+y^{n+1}}{2}\right),
 \label{eq:IMR}
\end{equation}
which in general involves the solution of a nonlinear equation. In some very specific cases, the equation above is invertible and one can solve for $y^{n+1}$ explicitly. For complicated systems of more variables such as equations (\ref{eq:geodesic3p1x}-\ref{eq:geodesic3p1u}) considered here, a common approach is to employ a Newton iterative scheme to address this task (see e.g. \citealt{press}).

This approach is clearly more computationally expensive, but it does bring the benefit of unconditional stability and symplecticity, provided that the iterative solution converges up to the prescribed tolerance. First integrals such as the associated Hamiltonian are not conserved exactly, but the error is bounded in time and of order $O(\Delta t^3)$ (\citealt{springel2014}). Moreover, the implicit nature of the scheme allows for larger $\Delta t$ without compromising stability. Hence, in the long term, these two features make a second-order implicit scheme such as the IMR superior to a higher-order explicit scheme such as the RK4.

\subsection{Energy-conserving scheme based on a Hamiltonian formulation}
\label{sec:ham}
Despite the generally good properties of symplectic schemes, in some very demanding cases the lack of exact energy conservation can be detrimental for the accuracy of the results. In such situations, the only solution is the use of a scheme that does conserve energy exactly. Such schemes have been widely studied and applied to systems characterized by a separable Hamiltonian (\citealt{fengqin}). However, for more general cases the construction of an exactly energy-conserving scheme can be complex. For the motion of charged particles in electromagnetic fields in special relativistic regimes, it has been shown that a slightly modified version of the IMR scheme is energy-conserving (\citealt{ripperda2018}). However, it is not straightforward to extend the same argument to our case of interest, given the difference in the governing equations and the much higher complexity of particle motion in general relativistic contexts. Here, we employ physical arguments in order to construct a second-order, Hamiltonian-preserving scheme suitable for the system of equations (\ref{eq:geodesic3p1x}-\ref{eq:geodesic3p1u}) from physical arguments.

The Hamiltonian for stationary metrics is defined as (see e.g. \citealt{gourgoulhon})
\begin{equation}
 H(\textbf{x},\textbf{u})=\alpha(\gamma^{jk} u_j u_k+\epsilon)^{1/2} - \beta^j u_j,
 \label{eq:hamiltonian}
\end{equation}
where $\textbf{x}=(x^1,x^2,x^3)$ and $\textbf{u}=(u_1,u_2,u_3)$. It is easy to see that $H$ represents the energy of a free particle traveling along a geodesic. Consider again the Killing vector, $K=\partial_t$, that characterizes any stationary spacetime. Killing's equation implies that the quantity
\begin{equation}
 K^\mu u_\mu=u_0
\end{equation}
is conserved along the particle trajectory. The zeroth component of the covariant four-momentum is
\begin{equation}
 u_0=-\alpha^2 u^0 +\beta^j u_j,
\end{equation}
and via the definition of $u^0=(\gamma^{jk} u_j u_k+\epsilon)^{1/2}/\alpha$, we retrieve $|u_0|=|H|$.

The result above suggests that any discrete integration scheme capable of conserving the Hamiltonian (\ref{eq:hamiltonian}) is also an energy-conserving scheme. Constructing such a scheme, however, is nontrivial. It is straightforward to derive the equations of motion by differentiating $H$, such that
\begin{equation}
 \frac{d x^i}{dt}=\frac{\partial H(\textbf{x},\textbf{u})}{\partial u_i}=\frac{\alpha \gamma^{ij} u_j}{(\gamma^{jk} u_j u_k+\epsilon)^{1/2}} -\beta^i,
\end{equation}
\begin{equation}
 \frac{d u_i}{dt}=-\frac{\partial H(\textbf{x},\textbf{u})}{\partial x^i}=-(\gamma^{jk} u_j u_k+\epsilon)^{1/2}\partial_i \alpha -\frac{1}{2}\frac{\alpha u_j u_k}{(\gamma^{jk} u_j u_k+\epsilon)^{1/2}}\partial_i \gamma^{jk} + u_j\partial_i \beta^j,
\end{equation}
which are precisely equations (\ref{eq:geodesic3p1x}-\ref{eq:geodesic3p1u}) above. For the equivalent discrete time integration scheme, one must ensure that the total variation of $H$ remains zero.

The key to achieving energy-conserving discretization is to consider how the variation of the Hamiltonian vanishes in the continuous case. Via the chain rule,
\begin{equation}
 \begin{aligned}
 \frac{d H(x^l,u_l)}{dt} & =\frac{\partial H(\textbf{x},\textbf{u})}{\partial x^i}\frac{d x^i}{dt}+\frac{\partial H(\textbf{x},\textbf{u})}{\partial u_i}\frac{d u_i}{dt} \\
 & = -\frac{du_i}{dt}\frac{dx^i}{dt}+\frac{dx^i}{dt}\frac{du_i}{dt} \\
 & =0.
 \end{aligned}
\end{equation}
The exact same argument can be extended to the discretized equations. Hence, one can infer that the discretized system
\begin{equation}
 \frac{\Delta x^i}{\Delta t}=\frac{\Delta H(\textbf{x},\textbf{u})}{\Delta u_i},
 \label{eq:deltax}
\end{equation}
\begin{equation}
 \frac{\Delta u_i}{\Delta t}=-\frac{\Delta H(\textbf{x},\textbf{u})}{\Delta x^i},
 \label{eq:deltau}
\end{equation}
will satisfy the above condition, leading to $\Delta H(\textbf{x},\textbf{u})/\Delta t=0$, provided that the discretized right-hand sides of equations (\ref{eq:deltax}) and (\ref{eq:deltau}) are defined such that this condition is fulfilled.

Numerical schemes derived from a discrete Hamiltonian have been applied extensively in many contexts, from molecular dynamics to the modeling of musical instruments (e.g. \citealt{tuckermanmartyna2000}; \citealt{chatziioannouvanwalstijn2015}). In this case, the Hamiltonian is in fact a function of six variables (three position components and three momentum components). It is clear from the equations above that, while the differentiation with respect to one variable is straightforward, it is not trivial to appropriately average each right-hand side of the equations with respect to the other variables:
\begin{equation}
 \frac{x^{i,n+1}-x^{i,n}}{\Delta t}=\frac{\overline{H}(\textbf{x},u_i^{n+1},u_l,u_m)-\overline{H}(\textbf{x},u_i^n,u_l,u_m)}{u_i^{n+1}-u_i^n},\qquad l,m\neq i,
\end{equation}
\begin{equation}
 \frac{u_i^{n+1}-u_i^{n}}{\Delta t}=-\frac{\overline{H}(x^{i,n+1},x^l,x^m,\textbf{u})-\overline{H}(x^{i,n},x^l,x^m,\textbf{u})}{x^{i,n+1}-x^{i,n}},\qquad l,m\neq i,
\end{equation}
where $\overline{H}$ indicates some average relative to the variables that are not affected by the differentiation. One such averaging for 2D systems can be found in \cite{fengqin}. A straightforward extension to 3D developed with the same philosophy can be found in Appendix \ref{app:ham}. There, we also discuss some peculiar aspects of the solution procedure.

The discrete equations above represent a new system of 6 nonlinear, coupled, first-order differential equations, that is more complex than the IMR scheme (\ref{eq:IMR}). In general, an iterative solution must be employed. The additional complexity comes at the benefit of attaining exact energy conservation regardless of the system parameters such as $\Delta t$, provided that the iteration is carried on up to convergence to machine precision.

A remarkable feature of this scheme is the absolute freedom in the definition of the Hamiltonian $H$. This implies that for systems characterized by a Hamiltonian different from that of equation ($\ref{eq:hamiltonian}$), the algorithm retains its energy (or in general, first integrals) conservation properties. Thus, the extension to more complicated physical situations becomes straightforward, provided that the corresponding Hamiltonian formulation is available.

\section{Tests}
\label{sec:tests} 
In this section, we test the methods described above with the aim of assessing both the accuracy of the results in a number of physically meaningful situations and the versatility of our implementation to handle generic spacetimes. Everywhere in the next subsections we employ geometrized units with $c=G=1$, such that time, mass, and distances are measured with the same units. The RK4 scheme, the IMR scheme, and the Hamiltonian scheme are tested and the results compared for standard spacetimes (Schwarzschild and Kerr). Note that for both metrics the geodesic equation can be solved analytically, see e.g. \cite{chandrasekhar1984}. This provides a powerful theoretical ground to test the accuracy of our implementation against known theoretical results. Although analytic solutions are available for these spacetimes (in terms of elliptic integrals), in ray-tracing algorithms for GRMHD frameworks they are rarely used (e.g. \citealt{chan2017}; \citealt{porth2017}), both due to the need of evaluating complicated expressions involving non-elementary functions (which makes the scheme error-prone) and to the poor versatility of the resulting algorithm. In this regard, using numerically calculated solutions is often faster and more flexible, e.g. when introducing perturbations to such standard spacetimes, for metrics that do not possess analytic solutions, and for non-analytic metrics defined on spatial grids.

\subsection{Tests in Schwarzschild spacetime}
The Schwarzschild solution to Einstein's equations describes a metric outside of a spherically symmetric body with a total mass $M$ in vacuum:
\begin{equation}
 ds^2=-\left(1-\frac{r_S}{r}\right)dt^2+\left(1-\frac{r_S}{r}\right)^{-1} dr^2+r^2d\theta^2+r^2\sin^2\theta d\varphi^2,
 \label{eq:schwarzschild}
\end{equation}
where $r_S=2M$ is the Schwarzschild radius. The metric presents a coordinate singularity at the event horizon at $r = r_S$ and a physical singularity at $r=0$. For $r \rightarrow \infty$ the Schwarzschild metric approaches Minkowski spacetime.

The metric possesses two conserved quantities: the energy, $E$, and a conserved angular momentum $L$ due to the symmetry with respect to rotations in $\varphi$. The motion of test particles in Schwarzschild spacetime restricted to the $\theta=\pi/2$ plane can be completely described by an effective potential (see e.g., \citealt{carrol}),
\begin{equation}
 V_{eff}(r)=\epsilon\left(1-\frac{r_S}{r}\right)+\frac{L^2}{r^2}-\frac{r_S L^2}{r^3},
 \label{eq:effpot}
\end{equation}
and the conserved energy $E$. Since in the plane $\theta=\pi/2$ the number of degrees of freedom is equal to the number of conserved quantities, the second-order geodesic equations (\ref{eq:geodesic}) are integrable. This feature is often used in geodesic integrators that take advantage of symmetries of the metric to simplify the equations to be solved (see e.g. \citealt{johanssen2010} or \citealt{johanssen2013}). In our implementation, however, there is no need to customize the integration for a specific metric, since the procedure is generally applicable to any 3+1 split spacetime.

The relativistic effective potential in equation (\ref{eq:effpot}) describes particle trajectories that cannot be represented by the corresponding Newtonian potential. The path followed by particles is ``open'' if the central object deviates the motion from a straight line, without keeping the particle on closed (periodic) orbits. Closed orbits, instead, correspond to the particle being trapped inside a characteristic potential well, determined by the conserved angular momentum $L=u_\varphi$. For massless particles, the potential well degenerates to a point, that identifies an unstable circular orbit at $r=3r_S/2$.

The Schwarzschild metric from equation (\ref{eq:schwarzschild}) is the simplest spherically symmetric solution to Einstein's equations of general relativity in static, vacuum spacetime. Despite its simplicity, it represents a reliable model for the theoretical study of distant, non-spinning massive objects that that are not directly observable. Hence, it is a widely used tool for the study of the properties of black holes, e.g. their gravitational lensing effects (\citealt{muller2008}). Simulations aimed at modeling the deflection of light rays caused by massive objects require the numerical integration of null geodesics. For massive particles the motion is described by timelike geodesics. For accretion flows around, e.g., neutron stars and black holes, modeling the motion of massive (often charged) particles is essential to gain insight in the microscopic dynamics of accretion disks and astrophysical outflows like jets and flares. 

\subsubsection{Deflection of light and ray tracing}
\label{sec:schwarzph}
Ray-tracing of light coming from very distant objects allows us to compare future observations with predictions from the current models. The usual ray-tracing approach consists of integrating the photon paths backwards, from the observer's position to the distant source. In the case of black holes, a recent burst of interest is directed towards observing the black hole shadow (\citealt{falcke2000}). The Event Horizon Telescope (EHT) project aims at providing, for the first time, images of Sgr A\mbox{*}, the black hole at the galactic center, and of the supermassive black hole in the center of M87 (\citealt{lobanov2017}; \citealt{akiyama2017}). Simulations of the expected shape of the black hole shadow will shed light on our understanding of the underlying physics. In tracing light rays coming from very distant objects, the accuracy of the computation is crucial in producing results that can be directly compared to measurements of the black hole shadow (see e.g. \citealt{straub2012} or \citealt{psaltis2016}).

As underlined in Section \ref{sec:rk4}, explicit methods such as the RK4 scheme introduce errors both in the computed variables (in this case, position and covariant velocity) and in the global conserved quantities (such as the energy). The longer the computation is carried on, the larger the accumulated energy errors can grow. In this Section, we analyze how energy errors can influence the results, and whether an energy-conserving method, that retains errors in $x^i$ and $u_i$ only, provides advantages versus a standard explicit method.

\begin{figure}[!h]
\centering
 \includegraphics[width=0.5\columnwidth]{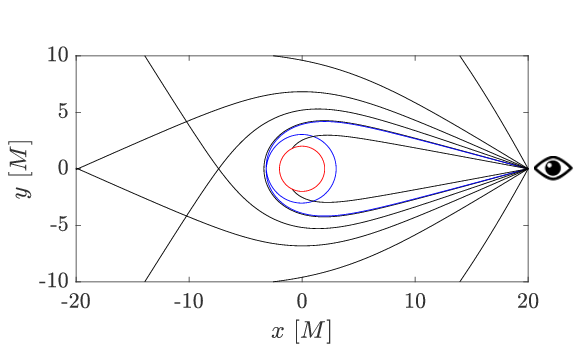}
 \caption{Deflection of light due to a Schwarzschild black hole. Light rays originating at the boundary of the box at $(20,0)$ under different angles are deflected, orbit or plunge into the compact object with its Schwarzschild radius indicated by the red circle. Secondary Einstein rings, circling the object once before arriving back at the departure point, are indicated in blue. Distances on both axes are measured in units of $M$.}
 \label{fig:phorbits}
\end{figure}

In our test, we initialize several light rays from the observer's position at $r=10r_S$ on the right-hand side of a central black hole of mass $M=1$ (central red circle in Figure \ref{fig:phorbits}). For the initialization setup, we refer to the detailed explanation in \cite{mullergrave2010}. Depending on the initial direction of propagation, the photons are strongly or weakly deflected from the straight path corresponding to Minkowski spacetime. Figure \ref{fig:phorbits} shows a few selected photon orbits. For specific choices of the initial direction the photon reaches a second observer located at the behind the compact object, or even returns to the first observer at the right-hand side, after circling the object a number of times that depends on the initial angle of propagation. In this way, Einstein rings of light are formed that provide the observer with an image of him or herself. Simulating this phenomenon requires high accuracy both in the initialization of the orbit and in the integration of the geodesic equation.

\begin{figure}[!h]
\centering
 \includegraphics[width=0.5\columnwidth]{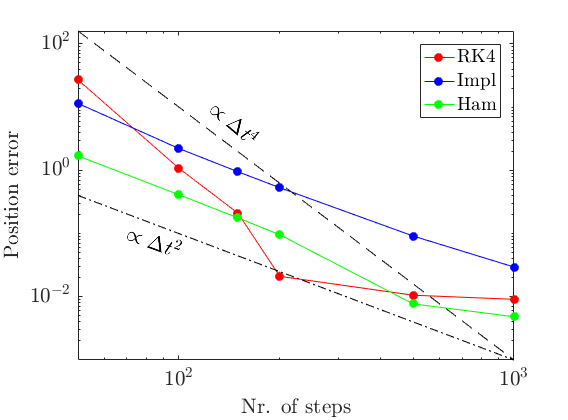}
 \caption{Representation of the error trends from the results listed in Table \ref{tab:ring2}. The second-order or fourth-order character of the integration schemes are highlighted by the decrease in the absolute error in the final position of the light ray. The Hamiltonian scheme (green line) performs better than the RK4 scheme (red line) for higher $\Delta t$. The IMR scheme (blue line) follows the same second-order trend as the Hamiltonian method, but it is characterized by larger errors. Above 200 integration steps, the error of the RK4 method starts to deviate from the reference fourth-order trend, due to higher accumulation of error as the number of integration steps increases.}
 \label{fig:ring2err}
\end{figure}

For a quantitative comparison, we refer to the setup in \cite{muller2008}: a light ray is initialized from $r_0=10 r_S$, with initial angle $\xi=0.24904964$ with respect to a straight line pointing from the right-hand observer to the compact object located at $r=0$. Under these initial conditions, the light ray forms a secondary Einstein ring and returns to the right-hand observer after one circle. The time it takes to form the ring can be computed analytically and corresponds to $\Delta t/r_S=50.396$ in coordinate time. Thus we can measure the final position $(r_f,\varphi_f)$ of the light ray at this time and check the deviation from the exact value, $(r_0=20,\varphi_0=0)$ (that is, the same position from which the light rays originated) by the usual formula, $|\Delta l|=\sqrt{r_f^2+r_0^2-2 r_f r_0 \cos(\varphi_0-\varphi_f)}$. Secondary Einstein rings approach the photon radius $3/2r_S$ and circle around it for one full period before reaching back the right-hand observer. Due to the proximity of the photon path to the event horizon, large energy errors can be expected if the integration is carried out with an explicit method. In this way, the path traveled by the light ray can severely deviate from the expected trajectory. Table \ref{tab:ring2} summarizes the results obtained by varying the number of integration steps along the path for the same final time, for each method. Figure \ref{fig:ring2err} shows the error trend, for each method, relative to the number of integration steps. The plot clearly shows the second-order (for the IMR and Hamiltonian) and fourth-order (for the RK4) character of the integration schemes adopted for this test.

\begin{table}[!h]
\centering
\begin{tabular}{|c|c|c|c|}
\hline
Nr. of steps & RK4 & Implicit & Hamiltonian \\ 
\hline 
50 & 26.97314 & 11.44041 & 1.68939 \\ 
\hline 
100 & 1.07701 & 2.21514 & 4.10968$\times 10^{-1}$ \\ 
\hline 
150 & 2.10550$\times 10^{-1}$ & 9.50488$\times 10^{-1}$ & 1.77281$\times 10^{-1}$ \\ 
\hline 
200 & 7.16948$\times 10^{-2}$ & 5.31338$\times 10^{-1}$ & 9.57071$\times 10^{-2}$ \\ 
\hline
500 & 1.05848$\times 10^{-2}$ & 9.13739$\times 10^{-2}$ & 7.72515$\times 10^{-3}$ \\ 
\hline
1000 & 9.12120$\times 10^{-3}$ & 2.95672$\times 10^{-2}$ & 4.83819$\times 10^{-3}$ \\ 
\hline
\end{tabular} 
\caption{Absolute error on the final position of a ray of light forming a secondary Einstein ring at the observer's position, integrated up to $t/r_S=50.396$, measured as $|\Delta l|=\sqrt{r_f^2+r_0^2-2r_f r_0 \cos(\varphi_0-\varphi_f)}$. The order of magnitude of the error introduced by the Hamiltonian scheme is the same as for the error introduced by the RK4 scheme, and in all cases, except for the case with 200 steps, the error for the Hamiltonian scheme is smaller.}
\label{tab:ring2}
\end{table}

The results show clearly that, for integrations performed with a relatively high time step, a second-order energy-conserving method is capable of retrieving the correct orbit better than a fourth-order explicit method. Decreasing $\Delta t$ decreases the error of the RK4 scheme at a faster rate than for the Hamiltonian scheme. When integrating with 1000 steps, the error starts to become comparable, since the order of the scaling of the RK4 error with the decreasing $\Delta t$ is higher than that of the Hamiltonian scheme (fourth-order and second-order, respectively). For this run, the energy error introduced by the RK4 scheme evidently dominates over the fourth-order accuracy of the method. In our simulations, we observe that the influence of the energy deviation on the resulting path is greater for a secondary ring than for a primary ring, where the photon turns back to the origin position right after passing around the object, without completing a full circle. The proximity of the trajectory to the event horizon at the Schwarzschild radius plays a role in the energy error introduced during the integration. A path that keeps the photon closer to the object for a longer time allows for larger energy errors to accumulate, worsening the results. Higher order Einstein rings, with photons circling the object more than once before returning to the initial position, represent a challenging problem for an explicit method. The Hamiltonian scheme used here, in contrast, conserves energy up to machine precision, and shows significantly smaller errors in the resulting photon path, performing better than the RK4 scheme especially for larger time steps. The implicit midpoint scheme performs worse than the other schemes in general, proving that symplecticity is a non-essential feature in this case for attaining higher accuracy in the results.

\subsubsection{Massive particle orbits}
\label{sec:schwarzp}
The accretion flow around compact objects consists of a gas of massive, charged particles, called a plasma. While global MHD simulations are necessary in order to characterize the macroscopic physics of accretion disks, tracing single massive particles allows for the investigation of acceleration phenomena at the microscopic level. For charged massive particles, the inclusion of the Lorentz force is necessary in order to study the related dynamics; here, we restrict our study to neutral particles. The simulation of timelike geodesics around compact objects can be modeled in similar fashion as the photon trajectories considered in the previous section. In Schwarzschild spacetime, particle orbits can be completely described in terms of conserved quantities, namely the particle energy $E$, and the angular momentum $L$. Additionally, the spherical symmetry in the metric allows for the restriction of any orbit on a plane of constant angle $\theta$. Bound orbits in Schwarzschild spacetime correspond to the particle moving inside a potential energy well, confined between the radii 
\begin{equation}
 r_{out,in}=\frac{L^2}{r_S}\pm L\sqrt{\frac{L^2}{r_S}-3}
\end{equation}
of the inner unstable circular orbit (subscript $in$, corresponding to the negative sign) and the outer stable circular orbit (subscript $out$, corresponding to the positive sign) allowed by the effective potential function (\ref{eq:effpot}). In fact, the relativistic precession of orbits is an effect of the particle climbing up and down the potential well while conserving its energy and angular momentum.

In our tests, we refer to the ``periodic table'' of closed orbits in Schwarzschild spacetime, presented in \cite{levinperezgiz2008}. There, a full description of the character of nontrivial orbits is given. Our aim is to verify how accurately the considered numerical methods describe the particle motion. In particular, we assess the ability of a scheme to keep the particle on a closed orbit for multiple cycles. Considering stable orbits ensures that deviation from the geodesic path is an effect solely linked to numerical errors. Accumulation of energy errors can lead the particle to climb up the potential well, until it escapes to infinity or falls into the black hole. Such effect is inevitable in explicit methods, meaning that no matter the order of the method, at some (ideally, very late) point in time, the particle is expected to escape the bound orbit. Hence, methods that keep the energy error bounded are capable of keeping a particle on orbit for a longer time than explicit methods (or in general methods with unbounded energy errors).

As an example, we consider a central black hole of mass $M=1$ and the orbit from the bottom-right panel of Figure 9 in \cite{levinperezgiz2008}. The conserved quantities for this case are $E=0.987649$ and $L=3.9$. The orbit roughly resembles a four-leaf clover that precesses slowly, closing up on itself after approximately 1000 cycles. In order to initialize the orbit, it is sufficient to note that, in our implementation, $u_\varphi=L$. Then we are free to set $u_r=u_\theta=0$ and $\theta=\pi/2,\varphi=0$ such that the particle starts its motion with a velocity purely in the $\varphi-$direction, from an unknown radius $r$ to be determined. As outlined in \cite{levinperezgiz2008}, the variation of $r$ in this case can be described by
\begin{equation}
 r^4 (u^r)^2=E^2 r^4-(r^2-r_S r)(r^2+L^2),
\end{equation}
and setting $u^r=0$ one can solve for the initial $r$ determined by the chosen $E$ and $L$.

\begin{figure}[!h]
\centering
\subfloat{\includegraphics[width=0.5\columnwidth]{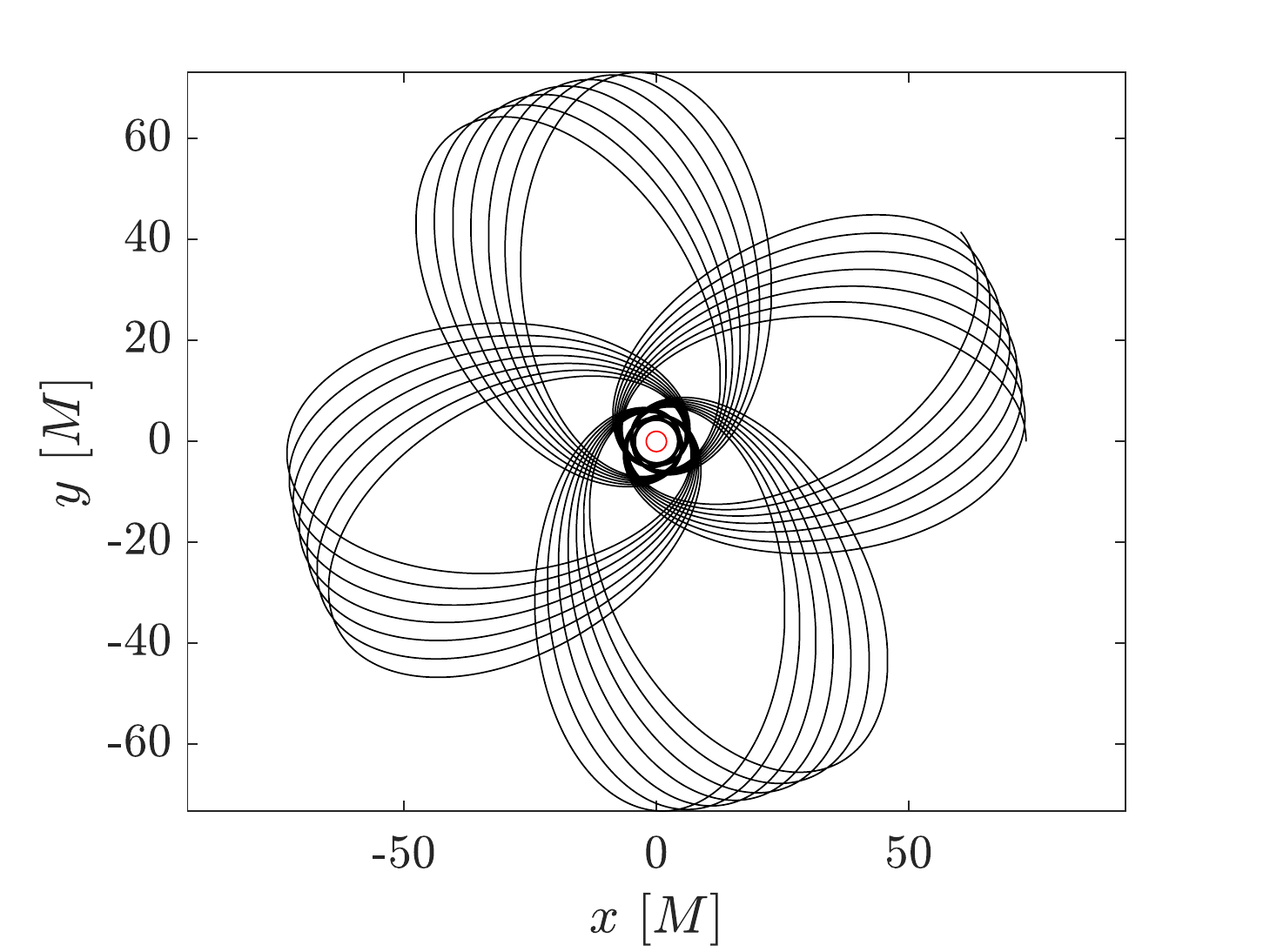}}
\subfloat{\includegraphics[width=0.5\columnwidth]{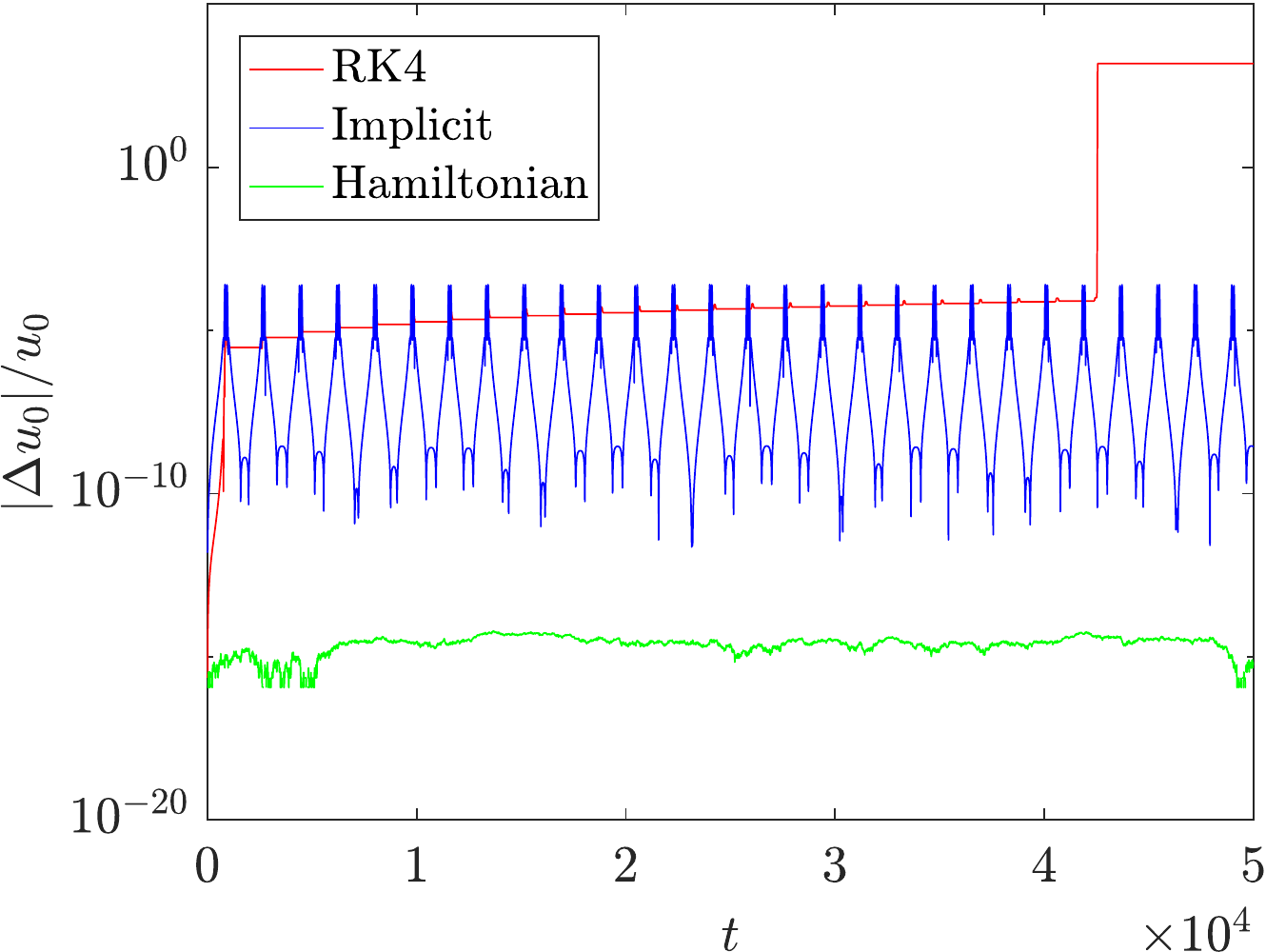}}
\caption{Simulation of the 4-leaf orbit with $E=0.987649$, $L=3.9$ from \cite{levinperezgiz2008}. Left-hand panel: a few precessions of a massive particle orbit (black line) around a compact object of Schwarzschild radius $r_S=2$ (red circle). The axes indicate scales in units of $M$. Right-hand panel: evolution of the energy error up to $t=5\times10^4$ when $\Delta t=10$. Around $t\approx4\times10^4$, accumulation of errors makes the RK4 solution (red line) blow up and lets the particle fly off from the orbit. This is clearly indicated by a sudden increase in the energy error. The error for the implicit scheme (blue line), although nonzero, is bounded and does not grow enough to let the particle escape. The same applies for the Hamiltonian scheme (green line), with errors of order machine precision, as expected.}
\label{fig:4leaf}
\end{figure}

Figure \ref{fig:4leaf} shows part of the orbit in the $r-\varphi$ plane (left-hand panel), as well as the evolution of the energy error in time up to $t=5\times10^4$ with $\Delta t=10$ (right-hand panel). Despite the large time step, the IMR scheme is capable of maintaining the energy error bounded, as expected from a symplectic method. The RK4 run, instead, fails after $t\approx 4\times10^4$, when accumulation of energy errors causes the particle to leave the orbit and fly away to infinity. The Hamiltonian scheme introduces even smaller energy errors, of order machine precision, as imposed by construction. Because of the absence of secular accumulation of energy errors, such a simulation could in principle continue indefinitely, without the particle ever escaping the bound orbit, regardless of the time-step size.

The results are expected, and in fact confirm that the implicit second-order IMR and Hamiltonian are superior to an explicit, although higher-order, scheme like the RK4, in the long run. In this case, the distinction between the two implicit schemes essentially reduces to the allowed time step size. The IMR scheme, being symplectic, conserves energy only to a certain degree, that scales with $\Delta t$. In case the integration step is too large, the achieved energy preservation may not be sufficient to constrain the particle motion to the theoretical bound orbit. The Hamiltonian method, instead, conserves energy regardless of the time step size, hence implying higher stability of the scheme at larger $\Delta t$. It should be noted that this test is highly idealized in excluding any external perturbation to the particle orbit, which eventually would lead to deviations from the nicely periodic closed path here represented. For cases in which integrating over a very large number of orbits is of interest, an explicit method is inferior to the two implicit methods considered here.

\subsection{Tests in Kerr spacetime}
\label{sec:kerr}
The Kerr solution to Einstein's equations describes a metric outside of a spherically symmetric body with a total mass $M$ and spin parameter $a$ in vacuum (\citealt{kerr1963}). In Boyer-Lindquist coordinates the metric is
\begin{equation}
ds^2 = -\left(1-\frac{2Mr}{\rho^2}\right)dt^2 - \frac{4Mra\sin^2\theta}{\rho^2}d\varphi dt + \frac{\rho^2}{\Delta}dr^2 + \rho^2 d \theta^2 + \left(r^2 + a^2 + \frac{2Mra^2 \sin^2 \theta}{\rho^2}\right)\sin^2 \theta d \varphi^2,
\label{eq:Kerr}
\end{equation}
with $\rho^2 \equiv r^2 + a^2 \cos^2 \theta$ and $\Delta \equiv r^2 - 2Mr + a^2$. This solution has coordinate singularities at the event horizons ($\Delta = 0$) and at the poles of the rotation axis. The singularities at the poles can be overcome by rewriting the metric in Cartesian Kerr-Schild coordinates.

The motion of test particles in the Kerr metric is not generally bound to a single plane of constant $\theta$. Hence it cannot be effectively determined by only one 1-dimensional potential function of the radius analogous to that characterizing the Schwarzschild solution. A description via such an effective potential is available, but only for orbits in the $\theta=0$ plane, see e.g. \cite{levinperezgiz2008}. Nevertheless, the metric retains enough symmetries to allow for the complete description of particle motion by the constants $E$, the conserved energy, $L$, the angular momentum, and $C$, Carter's constant. This also makes the geodesic equations integrable. However, as mentioned earlier, in our implementation there is no need to use symmetries in the metric to reduce the equations, since the algorithm applies to a generic 3+1 split metric. The parameter $a$, the black hole spin, causes objects at rest to move in the direction of rotation of the black hole, due to the so-called frame dragging, which is absent in the Schwarzschild case. Another feature of this metric is the presence of two event horizons, located at
\begin{equation}
 r_\pm=\frac{r_S\pm\sqrt{r_S^2-4a^2}}{2},
\end{equation}
where $r_S=2M$. The two event horizons coincide in case $a=M$, which describes a so-called extremal Kerr black hole, an unstable spinning compact object (\citealt{carrol}).

Simulating Kerr-type objects is of extreme importance, since most of the black holes that populate galaxies are expected to possess some spin, inherited from the originating stars. Future direct observations of such bodies, as well as global MHD simulation of accretion flows around them, must take the spin factor $a$ into account to provide correct results. In this context, both photon and massive particle trajectories are heavily influenced by the frame dragging effect, consistently differing from the zero-spin Schwarzschild approximation. In the next Section, we apply the presented numerical methods to the integration of both null and time-like trajectories. Note that, despite our choice of Boyer-Lindquist coordinates for simplicity, no extra difficulty arises in case one wants to adopt ``Cartesian'' Kerr-Schild coordinates. \cite{chan2017} show that the use of such a system of reference eliminates the singularities at the poles, and therefore it results more attractive for simulations of regions of spacetime around those points. The versatility of the 3+1 formulation of our framework allows for such change just by selecting the appropriate 3+1 metric functions.

\subsubsection{Unstable spherical photon orbits}
\label{sec:kerrph}
To test the strength of the numerical schemes, we simulate a set of unstable spherical photon orbits. For simplicity, we choose $a=M$ such that the inner and outer event horizons degenerate to one surface of constant radius $r_H=r_S/2$. \cite{teo2003} provides a full description of admitted spherical orbits restricted between the radii
\begin{equation}
 r_{r,p}=2M\left\{1+\cos\left[\frac{2}{3}\cos^{-1}\left(\pm\frac{|a|}{M}\right)\right]\right\},
\end{equation}
of retrograde (subscript $r$) and prograde (subscript $p$) circular equatorial orbits, respectively. Such orbits are fully characterized by the constants of motion, namely the orbit radius $r_0$ (in units of $M$), a measure of the angular momentum $\Phi=L/E$ (in units of $M$), and a measure of Carter's constant $\mathcal{Q}=C/E^2$ (in units of $M^2$). All such orbits are unstable, meaning that perturbations introduced in the parameters are amplified in time until the photon deviates from the orbital motion and flies away. \cite{chan2017} have studied a number of unstable photon orbits both in Boyer-Lindquist and in Cartesian Kerr-Schild coordinates by means of the RK4 integration scheme. An explicit scheme naturally introduces energy errors at each time step, as well as errors in the position and velocity. The Hamiltonian energy-conserving scheme eliminates energy errors, resulting in a more stable orbital motion. Here, we aim at evaluating how this feature impacts the results in terms of the capability of keeping the photon on the correct orbit.

We consider several orbits taken both from \cite{teo2003} and \cite{chan2017}, and summarized in Table \ref{tab:orbitspar}, around a central black hole of mass $M=1$. The initialization of each orbit is done according to the description provided in \cite{mullergrave2010}. Each orbit starts at the equator, $\theta=\pi/2$, at $\varphi=0$, with an initial velocity pointing southwards. The deviation in the radial position $r$ from the constant value characterizing each orbit grows exponentially in the beginning (\citealt{chan2017}), and can be monitored in order to measure the performance of each scheme. For each orbit, we integrate until $t=100$ with several values of $\Delta t$.

\begin{table}[!h]
\centering
\begin{tabular}{|c|c|c|c|c|c|c|}
\hline
Orbit name & $\Phi$ & $r_0$ & $\mathcal{Q}$ \\ 
\hline 
A & -1 & $1+\sqrt{3}$ & $12+8\sqrt{3}$ \\ 
\hline
B & -6 & $1+2\sqrt{2}$ & $-13+16\sqrt{2}$ \\ 
\hline
C & 1 & 2 & 16 \\ 
\hline
D & 1.36 & 1.8 & 12.8304 \\ 
\hline
\end{tabular} 
\caption{Parameters for the unstable spherical photon orbits considered in Section \ref{sec:kerrph}.}
\label{tab:orbitspar}
\end{table}

\begin{figure}[!h]
\centering
\subfloat[Orbit A.]{\includegraphics[width=0.5\columnwidth]{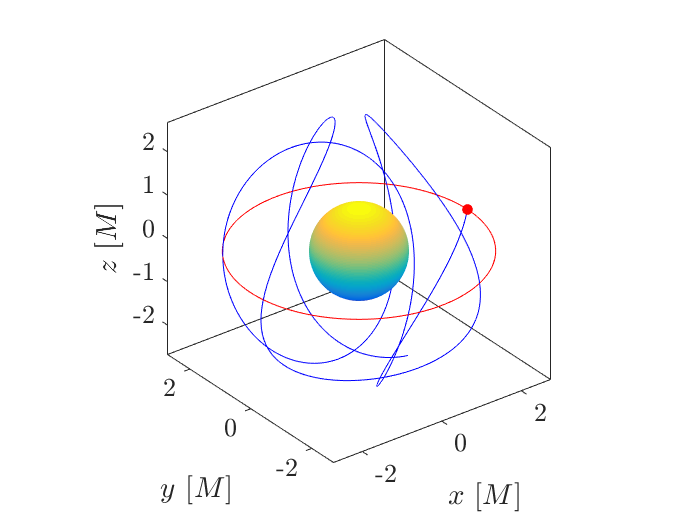}}
\subfloat[Orbit C.]{\includegraphics[width=0.5\columnwidth]{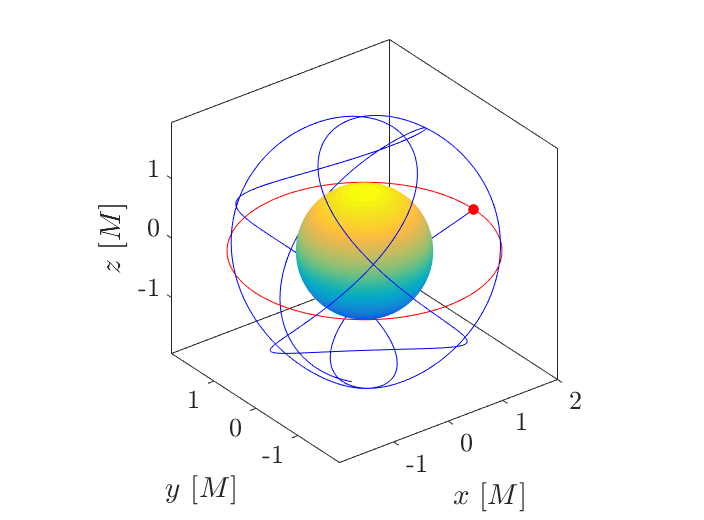}}
\caption{Representation in three-dimensional space of the unstable spherical photon orbits A (left-hand panel) and C (right-hand panel). The event horizon at $\Delta=0$ is indicated by a sphere of radius $r_+=r_-$. The equatorial red circle indicates the constant radius $r_0$ characterizing each orbit. The starting point of the orbits is marked by a red dot.}
\label{fig:orbitBE}
\end{figure}

Figure \ref{fig:orbitBE} is a representation of the simulated orbits A (left-hand panel) and C (right-hand panel) in three-dimensional space. Here, the event horizon is represented as a sphere of radius $r_H$. The sense of rotation of the central object is from left to right. The red circle of radius $r_0$ marks the value of $r$ on which, in absence of perturbations, the spherical orbit should remain. Orbit A (left-hand panel) is a retrograde orbit, where the angular momentum $L$ is high enough to overcome frame dragging and the photon precesses in the direction opposite to the rotation of the central object. Orbit C (right-hand panel), on the other hand, is a prograde orbit. Both orbits start from the equator ($\theta=\pi/2$) and head southwards.

\begin{figure}[!h]
\centering
\subfloat[Orbit A.]{\includegraphics[width=0.5\columnwidth]{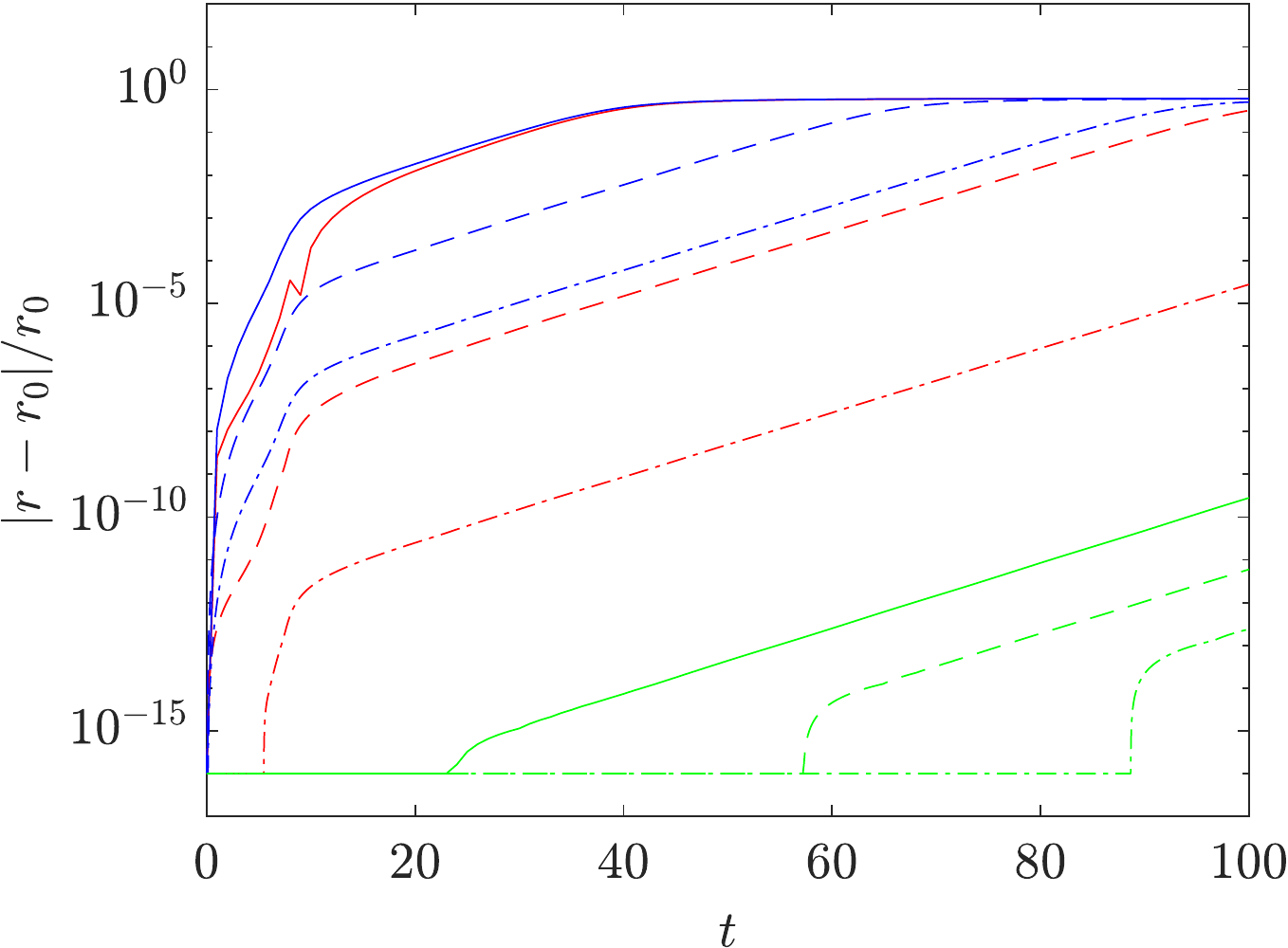}}
\subfloat[Orbit C.]{\includegraphics[width=0.5\columnwidth]{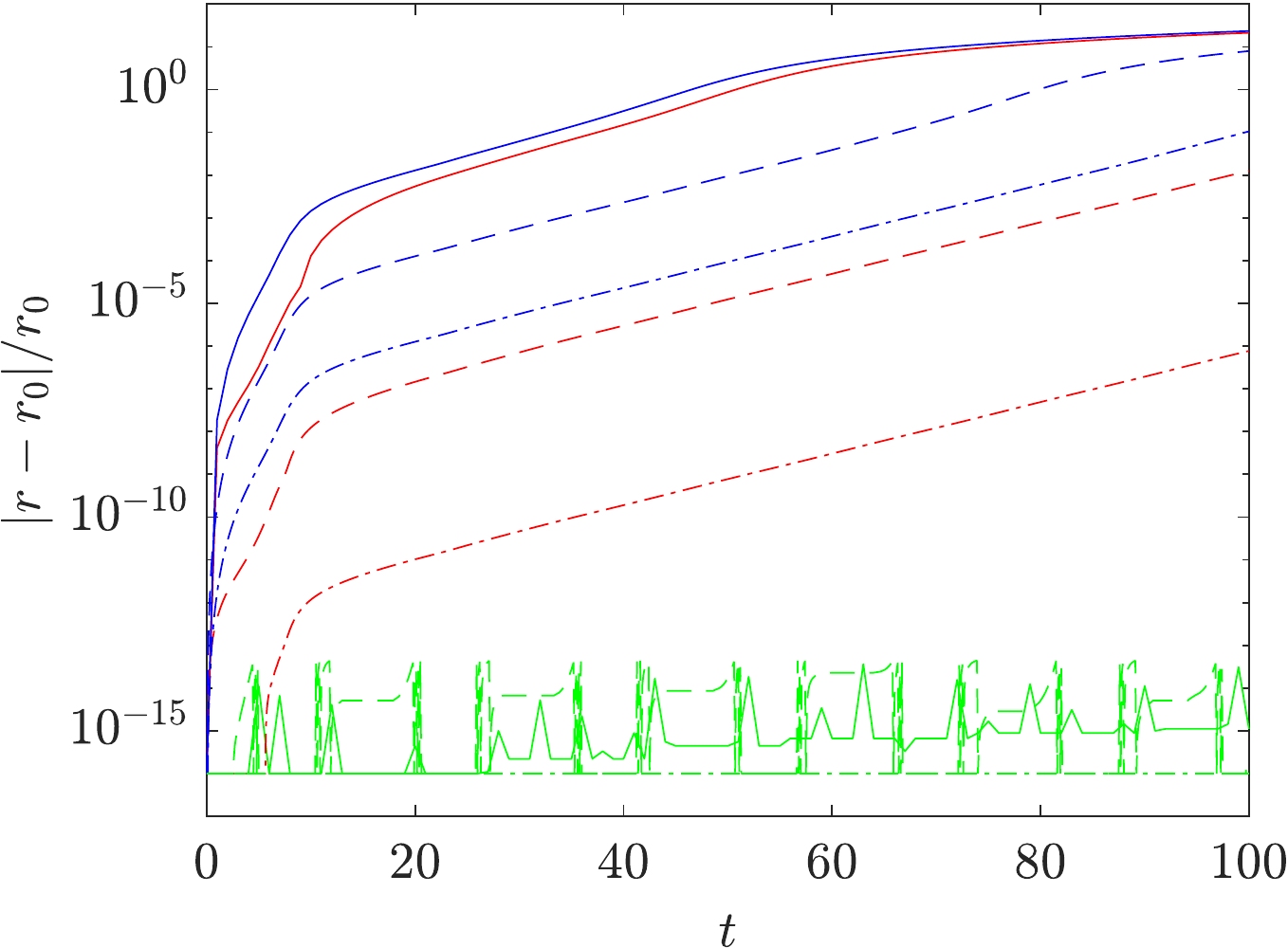}}
\caption{Time evolution of the relative error on the radius for orbits A (left-hand panel) and C (right-hand panel). Each simulation is run until $t=100$ with $\Delta t=1$ (solid lines), $\Delta t=0.1$ (dashed lines), $\Delta t=0.01$ (dash-dotted lines) with the RK4 (red lines), implicit (blue lines) and Hamiltonian (green lines) schemes.}
\label{fig:orbitBEerr}
\end{figure}

Figure \ref{fig:orbitBEerr} shows the evolution in time of the relative error on the radius $|r-r_0|/r_0$, for orbits A (left-hand panel) and C (right-hand panel) and for several values of $\Delta t$ for each method. Due to numerical errors, the growth of the error undergoes an exponential phase, in which the photon is still bounded to the orbit but progressively deviates from it. At the end of the exponential phase, the photon is eventually released and flies off. In general, the growth of the error is reduced by reducing the time step from $\Delta t=1$ (solid lines) to $\Delta t=0.1$ (dashed lines), and further to $\Delta t=0.01$ (dash-dotted lines), for all methods. The magnitude of the error for the RK4 scheme (red lines) reduces greatly when reducing the time step, as expected for a fourth-order method. The error from the implicit method (blue lines) shows similar behavior, though the error reduction is weaker, due to the method being second-order. Remarkably, the error from the Hamiltonian method (green lines) is orders of magnitude smaller than the error from the other methods, for all values of $\Delta t$. The energy-conserving properties of the algorithm maintain the photon on the orbit for much longer, delaying the beginning of the linear growth to much later with respect to the moment it starts in the case of the other two methods. The photon is still eventually escaping, due to accumulation of second-order errors, though the absence of energy errors greatly improves the stability of the orbit. For orbit E (right-hand panel), the linear growth of the error is not even observed within the simulated time, and the relative error on the position remains at machine precision. 

Table \ref{tab:kerrpherr} summarizes the results for selected orbits by listing the relative error on the radius at the last time step. The error for the Hamiltonian method is always orders of magnitude smaller than that from the RK4 runs. The error for the implicit method is always greater than that of the RK4. Errors of magnitude $\sim 1$ or larger should be regarded as a signature of the photon having left the orbit at the moment of evaluating the error. In all cases, the Hamiltonian scheme keeps the photon on the correct orbit within the simulated time, even for large time steps. The RK4 scheme needs much smaller time steps to prevent the photon from flying off, and the resulting error is still larger than that obtained with the Hamiltonian scheme. Orbit B shows the remarkable feature that the error obtained with the Hamiltonian scheme and $\Delta t=0.1$ is smaller than that for the $\Delta t=0.01$ run. This is a signature of the trade-off threshold between gain in accuracy and accumulation of error that characterizes the time step reduction. The larger number of time steps of the $\Delta t=0.01$ run accumulates larger total second-order errors, overcoming the gain in accuracy due to the reduction from $\Delta t=0.1$. This is an expected effect, that can be overcome solely by an increase of the order of the method.

\begin{table}[!h]
\centering
\begin{tabular}{|c|c|c|c|c|c|}
\hline
Method & $\Delta t$ &  Orbit A & Orbit B & Orbit C & Orbit D \\ 
\hline 
RK4 & $1$ & 0.620 & 1.613$\times 10^{-2}$ & 21.350 & 13.290 \\ 
\hline
RK4 & $0.1$ & 0.324 & 2.090$\times 10^{-4}$ & 1.263$\times 10^{-1}$ & 7.084$\times 10^{-4}$ \\ 
\hline 
RK4 & $0.01$ & 2.767$\times 10^{-5}$ & 0 & 7.722$\times 10^{-7}$ & 5.414$\times 10^{-8}$ \\ 
\hline 
IMR & $1$ & 0.621 & 0.728 & 23.450 & 23.080 \\ 
\hline
IMR & $0.1$ & 0.608 & 0.718 & 7.978 & 4.280 \\ 
\hline 
IMR & $0.01$ & 0.513 & 0.626 & 1.057$\times 10^{-1}$ & 1.794$\times 10^{-2}$ \\ 
\hline 
Hamiltonian & $1$ & 2.817$\times 10^{-10}$ & 7.995$\times 10^{-12}$ & 1.11$\times 10^{-15}$ & 1.494$\times 10^{-13}$ \\ 
\hline
Hamiltonian & $0.1$ & 5.972$\times 10^{-12}$ & 2.900$\times 10^{-14}$ & 0 & 9.659$\times 10^{-14}$ \\ 
\hline 
Hamiltonian & $0.01$ & 2.445$\times 10^{-13}$ & 7.956$\times 10^{-13}$ & 0 & 6.131$\times 10^{-14}$ \\ 
\hline 
\end{tabular} 
\caption{Relative error on the radius for several unstable spherical photon orbits at the final time $t=100$.}
\label{tab:kerrpherr}
\end{table}

\begin{figure}[!h]
\centering
\subfloat[Orbit A.]{\includegraphics[width=0.5\columnwidth]{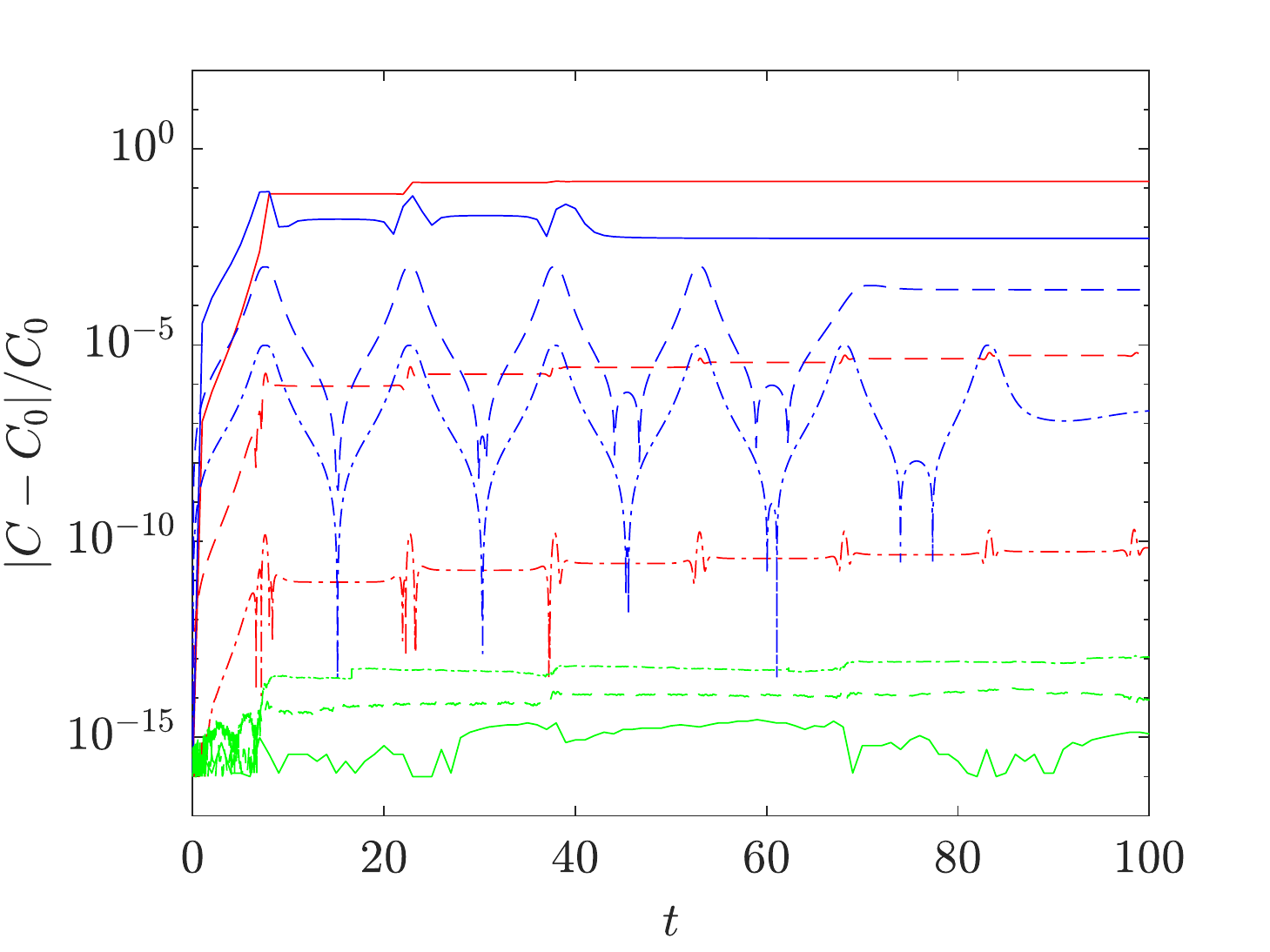}}
\subfloat[Orbit C.]{\includegraphics[width=0.5\columnwidth]{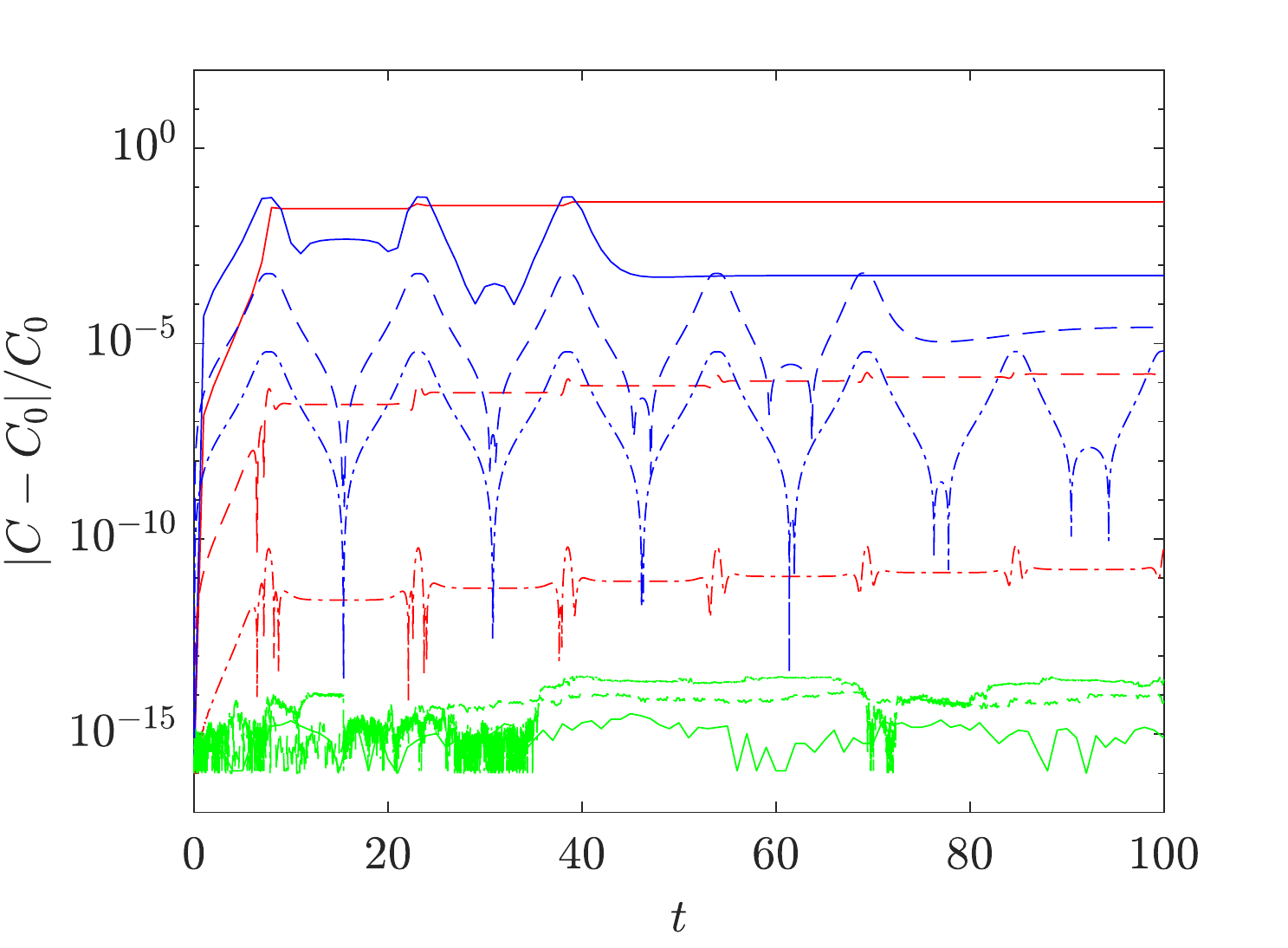}}
\caption{Time evolution of the relative error on the Carter constant $C_0$ for orbits A (left-hand panel) and C (right-hand panel). The same color schemes of Figure \ref{fig:orbitBECerr} are used here, with $\Delta t=1$ (solid lines), $\Delta t=0.1$ (dashed lines), $\Delta t=0.01$ (dash-dotted lines) for the RK4 (red lines), implicit (blue lines), and Hamiltonian (green lines) schemes.}
\label{fig:orbitBECerr}
\end{figure}

As a final test, we check the conservation of the Carter constant $C$, expressed as
\begin{equation}
 C = u_\theta+\cos^2\theta\left(a^2(\epsilon-E^2)+\frac{L^2}{\sin^2\theta}\right),
\end{equation}
where $L$ and $E$ are the angular momentum and energy, and $\epsilon=0$ for massless particles. Note that checking for the conservation of $L$ is not meaningful in this case, since the geodesic equation in Boyer-Lindquist coordinates naturally reduces to $du_\phi/dt=dL/dt=0$, and therefore the angular momentum is automatically conserved due to our choice of coordinates. Figure \ref{fig:orbitBECerr} shows the time history of the relative error with respect to the initial value $C_0$ of the Carter constant for orbits A (left-hand panel) and C (right-hand panel) analyzed above, with the same decreasing values of $\Delta t$ for all three numerical methods. The plots show clearly that the conservation of $C_0$ is achieved to machine precision for all $\Delta t$ by the Hamiltonian method, while the RK4 and IMR schemes retain a nonzero error in all cases. Therefore, for this case, the energy-conserving character of the Hamiltonian scheme results in the conservation of the other invariants of the motion as well.

\subsubsection{Massive particle orbits}
\label{sec:kerrp}

For completeness, we include an example of massive particle orbits around a Kerr black hole. We refer again to \cite{levinperezgiz2008} (Figure 15, third panel from the left in the second row) for initializing a periodic orbit around a spinning object with $a=0.995M$ and $M=1$. Similarly to the corresponding case in Schwarzschild spacetime (see Section \ref{sec:schwarzp}), we set $u_\varphi=L$, and thus we are free to choose $u_r=u_\theta=0$ and $\theta=\pi/2$, $\varphi=0$, such that the particle is set in motion in the $r-\varphi$ plane with a purely azimuthal velocity. The unknown initial radius is given by the equation
\begin{equation}
 \rho u^r=\sqrt{[E(r^2+a^2)-aL]^2-\Delta[r^2+(L-aE)^2]},
\end{equation}
describing the variation of $r$, with $E=0.920250$ and $L=2$ for the chosen orbit. Setting again $u^r=0$ and solving for $r$ gives the necessary initial condition.

\begin{figure}[!h]
\centering
\includegraphics[scale=.5]{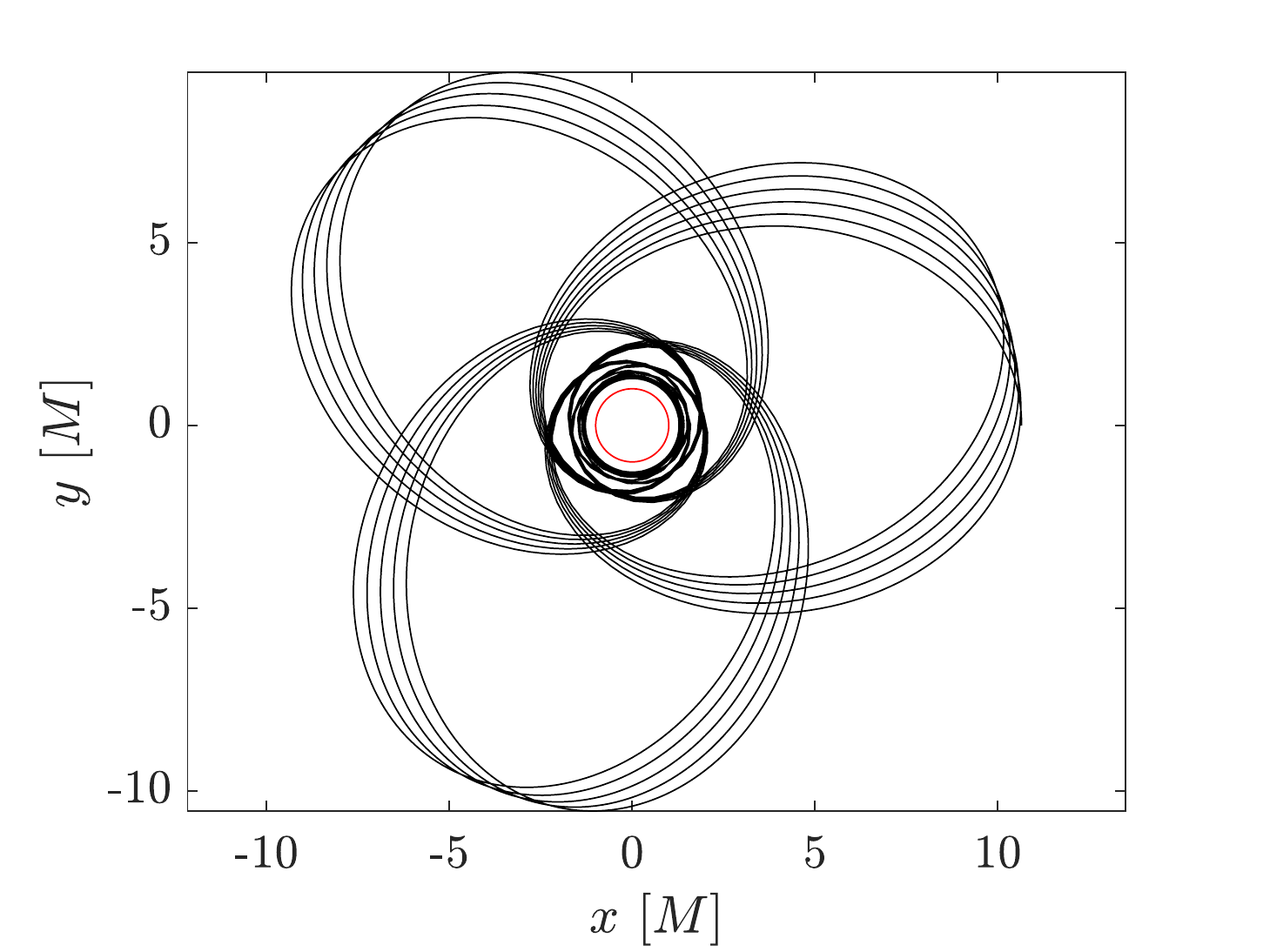}
\caption{Simulation of a few precessions of the 3-leaf orbit in Kerr spacetime, with $E=0.920250$ and $L=2$ from \cite{levinperezgiz2008}. The precessing orbit (black line) winds up around the central object with external event horizon of radius $r_+$ (red circle). The axes indicate scales in units of $M$.}
\label{fig:orbitp}
\end{figure}

A few precessions of the simulated orbit are shown in Figure \ref{fig:orbitp}. The analysis of the energy errors is completely analogous to the Schwarzschild case (see Figure \ref{fig:4leaf}), with the error in the RK4 scheme accumulating unboundedly until the particle escapes to infinity. The IMR scheme and the Hamiltonian scheme, on the other hand, are capable of keeping the particle on the expected trajectory indefinitely.

\section{Applications in other spacetimes}
\label{sec:applications}

In this section we present applications of our geodesics integrator that are relevant for future global simulations in more exotic spacetimes. Regardless of the metric characterizing each application, it is important that the solution procedure remains the same, in order to preserve the flexibility of the framework. Although we run each test with all the presented integrators, in order to check the convergence of the results, we are mainly interested in testing the versatility of the adopted generalized formulation rather than the accuracy of the solution.

\subsection{Morris-Thorne wormhole}
\label{sec:MT}
The Morris-Thorne solution to Einstein's equations describes a simple type of wormhole, representing the quantum foam that connects two distinct regions of spacetime (\citealt{morristhorne1988}). The metric in spherical coordinates is written
\begin{equation}
ds^2 = -dt^2 + dl^2 +(b_0^2+l^2)(d\theta^2 + \sin^2 \theta d \varphi^2),
\label{eq:morristhorne}
\end{equation}
where $b_0$ is the size of the wormholes throat and $l$ is the proper radial coordinate. 
A Morris-Thorne wormhole is traversable, in the sense that it is possible to travel from one side of the wormhole to the other and back. The observer's view through the wormhole is visualized by null geodesics. Due to spherical symmetry of the static metric, we can restrict to the two-dimensional hyperplane defined by $\theta=\pi/2$. This can be embedded in three-dimensional Euclidian space by the embedding function
\begin{equation}
z(r) = \pm b_0 \ln\left[\frac{r}{b_0} + \sqrt{\left(\frac{r}{b_0}\right)^2 -1}\right],
\label{eq:embeddingfunction}
\end{equation}
where $r^2 = b_0^2 + l^2$ (\citealt{mullergrave2010}).

\begin{figure}[!h]
\centering
\includegraphics[scale=.5]{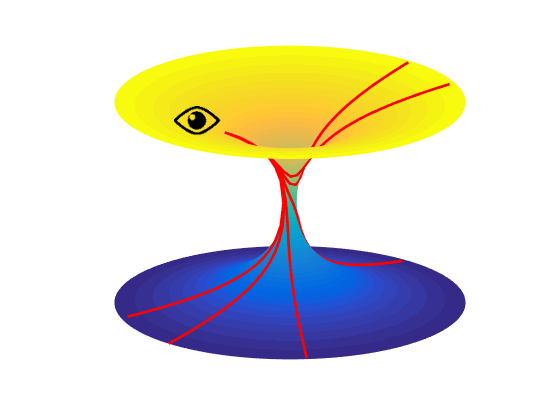}
\caption{Light rays (red line) originating from the observer's position at $l=10$ are deflected by a Morris-Thorne wormhole with throat size $b_0=1$. The light rays can remain in the same region of spacetime or travel through the wormhole to reach the other end, e.g. a causally separated region of the same universe.}
\label{fig:MT}
\end{figure}

To exemplify the capability of our implementation to handle generic spacetimes, we initialize several light rays from an observer's position in the vicinity of a Morris-Thorne wormhole of throat size $b_0=1$. The initialization is done according to \cite{mullergrave2010}. Depending on the initial direction of propagation, the photons can be deflected by an effect similar to that observed in a Schwarzschild metric, or travel through the wormhole to reach a different region of spacetime. Several photon paths are shown in Figure \ref{fig:MT}. When running the test with the three numerical integrators introduced in the previous sections, we find no significant differences in the resulting photon paths.

\subsection{Extremal Reissner-Nordstr\"{o}m dihole}
\label{sec:RN}
We now consider black holes with electric and magnetic charge, that are of particular interest for plasma physics. They are found by solving the Einstein equations with an electromagnetic radiation source. The Reissner-Nordstr\"{o}m metric describes a spherically symmetric mass distribution with total mass $M$ and charge $Q$:

\begin{equation}
ds^2 = \Delta (r) dt^2 + \frac{1}{\Delta(r)}dr^2 + r^2d\theta^2+r^2\sin^2\theta d\varphi^2,
\label{eq:RN}
\end{equation}
with 
\begin{equation}
\Delta(r) \equiv 1 - \frac{2GM}{r} + \frac{G Q^2}{r^2}.
\label{eq:deltaRN}
\end{equation}
Like in the Schwarzschild solution, $r=0$ describes a physical singularity, where the curvature blows up. When $g_{tt} = \Delta(r) = 0$, there are additional coordinate singularities at 
\begin{equation}
r_{\pm} = GM \pm \sqrt{GM^2 - GQ^2},
\label{eq:singularitiesRN}
\end{equation}
with $r = r_{\pm}$ corresponding to the event horizons. In the case $GM^2 = Q^2$, those two horizons coincide exactly. Therefore the metric is called extremal in this case. This solution is unlikely to occur in nature as it requires an enormous amount of charge. However, it is an ideal test case for numerical integrators, since exact solutions are known. \cite{mullerfrauendiener2011} provide a number of interesting orbits for a specific multi-black hole solution in the extremal Reissner-Nordstr\"{o}m metric, discussed in \cite{chandrasekhar1989}. A spacetime describing multiple Reissner-Nordstr\"{o}m black holes admits static solutions where the electric repulsion between the objects compensates exactly for the gravitational attraction, such that the resulting metric does not change in time. A dihole metric of this type reads, in Cartesian coordinates,
\begin{equation}
 ds^2=-\frac{1}{U^2}dt^2+U^2(dx^2+dy^2+dz^2),
 \label{eq:RNmetric}
\end{equation}
where $U=1+M_1/r_1+M_2/r_2$. Here, $M_1$ and $M_2$ are the masses of the two black holes, $r_1=\sqrt{x^2+y^2+(z-1)^2}$, and $r_2=\sqrt{x^2+y^2+(z+1)^2}$. The extremal dihole metric does not possess as many conserved quantities as degrees of freedom, hence it is non-integrable (\citealt{contopoulos1990}). Nevertheless, particle orbits in this spacetime can be classified and studied numerically.

\begin{figure}[!h]
\centering
\subfloat{\includegraphics[scale=.5]{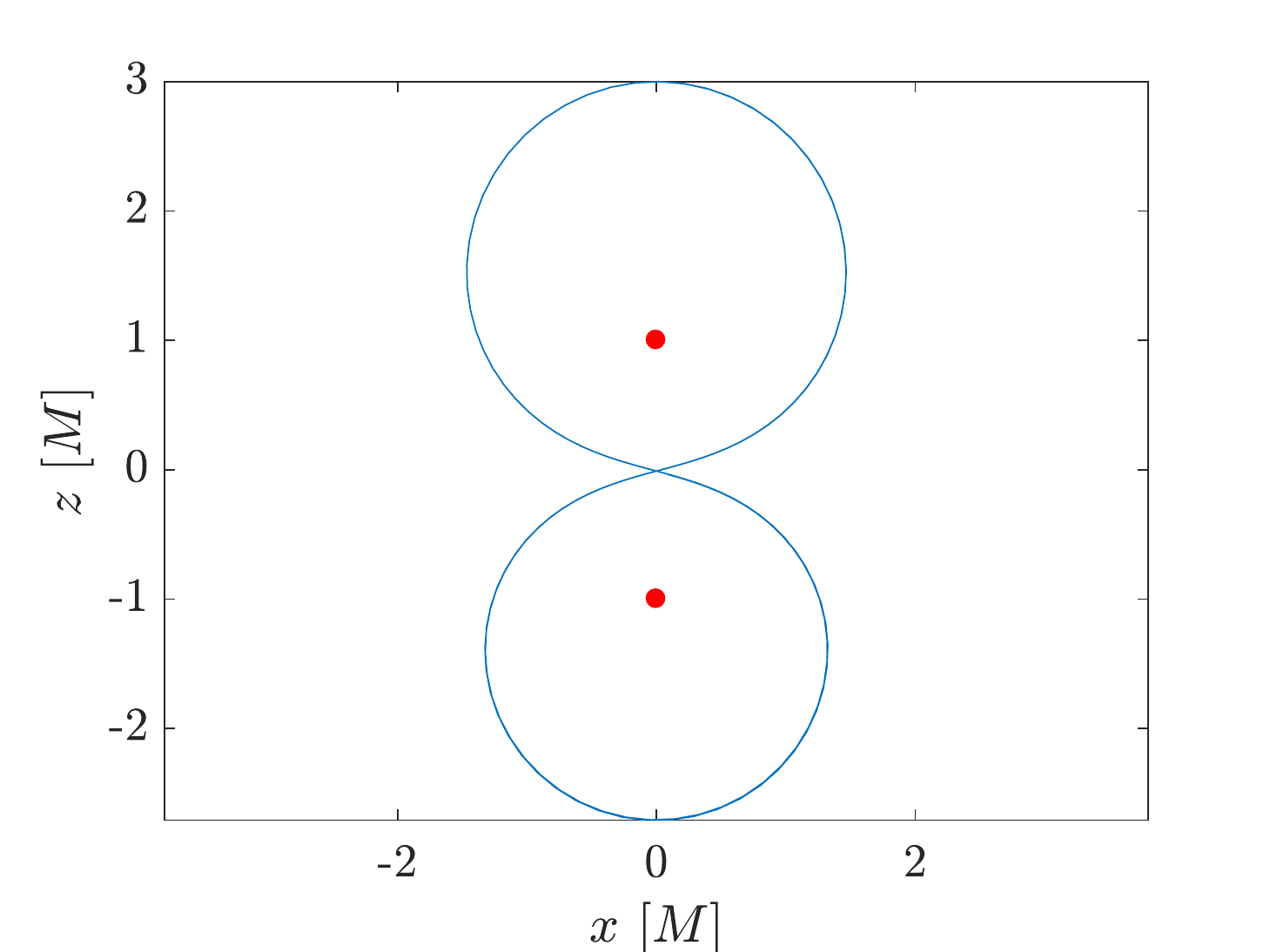}}
\subfloat{\includegraphics[scale=.5]{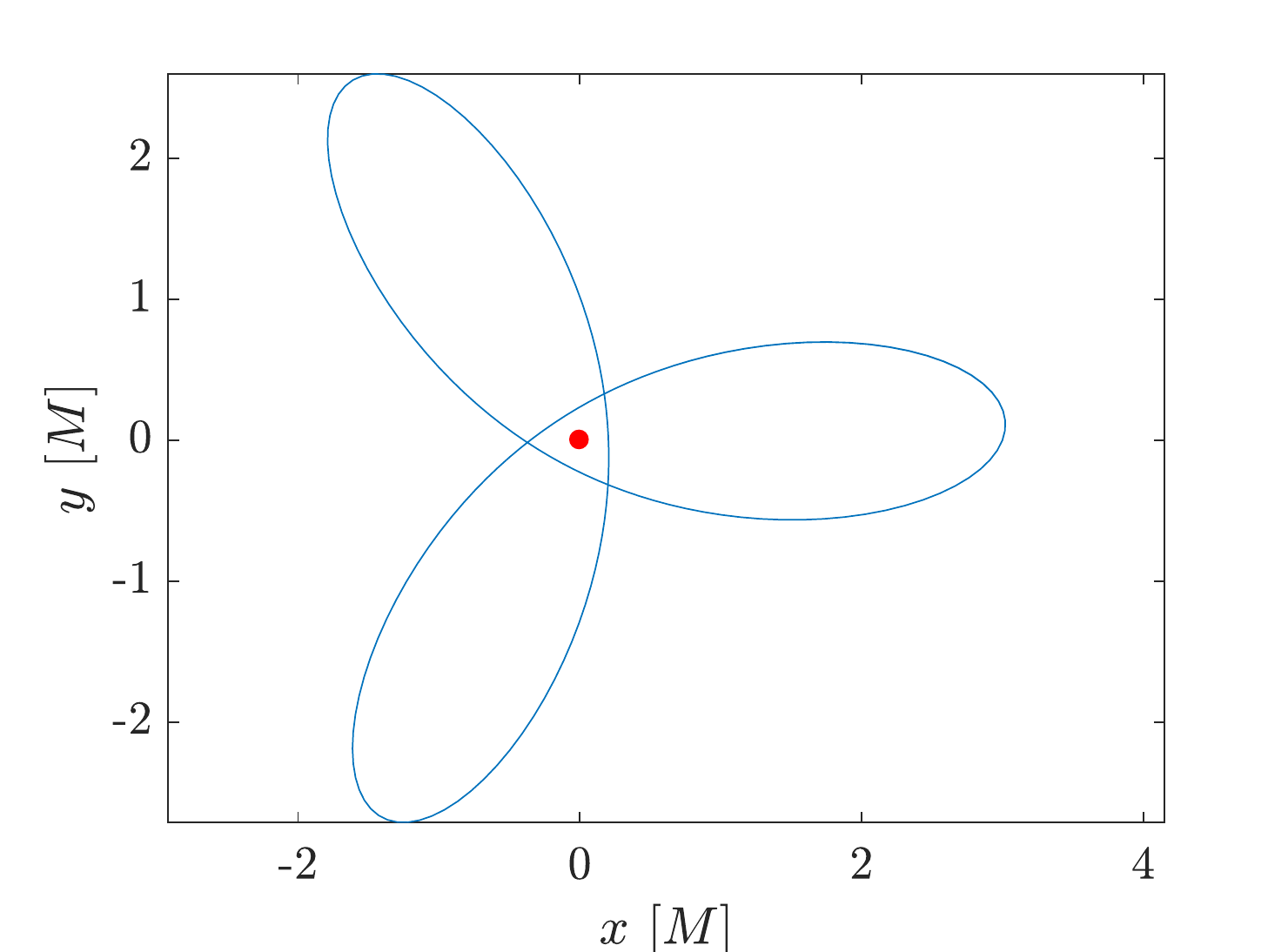}}
\caption{Simulation of closed massive particle orbits in the Reissner-Nordstr\"{o}m dihole metric (\ref{eq:RNmetric}). The position of the black holes is marked with red dots. Left-hand panel: a closed orbit in the $x-z$ plane. Right-hand panel: a three-leaf closed orbit in the $x-y$ plane.}
\label{fig:extrRN}
\end{figure}

For this test, we set $M_1=M_2=1$ and simulate closed particle orbits in this spacetime. Note that the metric (\ref{eq:RNmetric}) is such that the two black holes are located at $x=y=0,z=\pm1$. Additionally, it presents singularities only at the black holes locations, $r_1=0$ and $r_2=0$. We adopt the setups detailed in \cite{mullerfrauendiener2011} to initialize two periodic timelike geodesics. In both cases, the orbit starts from an initial position $\textbf{x}_0=(x_0,y_0,z_0)$ with 3-velocity $\textbf{u}_0=v\Gamma(\cos{\xi},\sin{\xi},0)/U(\textbf{x}_0)$, where $\Gamma=1/\sqrt{1-v^2}$. The two orbits are shown in Figure \ref{fig:extrRN}. The left-hand panel shows an orbit in the $x-z$ plane, starting from $x_0=y_0=0,z_0=3$ with $v=0.591943$, $\xi=0$. The right-hand panel is a three-leaf orbit in the $x-y$ plane, starting from $x_0=3,y_0=z_0=0$ with $v=0.2$, $\xi=1.2088$. The position of the black holes is indicated with red dots. The simulation of these orbits is quite challenging, since the presence of multiple black holes creates more than one pole of attraction for the path of the particles. In our runs, we find that the RK4 scheme performs better than the other schemes in reproducing the correct orbits. This can be explained by considering the order of the methods. 

The absence of event horizons allows an explicit method to introduce far smaller energy errors than in the case of e.g. the Schwarzschild metric, where large energy errors for the RK4 scheme arise for a particle orbiting close to the event horizon. Therefore, in this case, errors linked to the order of the method, resulting in inaccurate orbits, are predominant over energy errors. Considering that the presence of multiple black holes increases the precision necessary to keep a particle on the right path even further, we conclude that a higher-order method, though explicit, can perform better with respect to second-order implicit methods. In this case the RK4 scheme is a good choice for simulating such closed orbits, although the same does not necessarily hold for different paths passing close to the singularities.

\subsection{Perturbations to general relativistic spacetimes}
\label{sec:pertschwarz}
The versatility of our implementation allows for further exploration of nonstandard spacetimes. Motivated by the work by \cite{giddingspsaltis2016}, we simulate photon trajectories in a perturbed Schwarzschild metric, to assess the effects of deviations from standard metrics in general relativity. Here, we choose to follow the description of general perturbations provided by \cite{reggewheeler1957}. In their work, a perturbed metric $\tilde{g}_{\mu\nu}=g_{\mu\nu}+h_{\mu\nu}$ is determined by, e.g. the simplified ``even'' perturbation
\begin{equation}
 h_{\mu\nu}=
 \begin{pmatrix}
  \left(1-\frac{r_S}{r}\right)F(r)Y_L^M & 0 & 0 & 0 \\
  0 & \left(1-\frac{r_S}{r}\right)F(r)Y_L^M & 0 & 0 \\
  0 & 0 & r^2 K(r) Y_L^M & 0 \\
  0 & 0 & 0 & r^2 K(r) Y_L^M \sin^2\theta
 \end{pmatrix},
 \label{eq:schwarzpert}
\end{equation}
where $ Y_L^M $ is the scalar spherical harmonic function and $F(r),K(r)$ are functions determined by the modes $L,M$ of the perturbation, e.g. plane waves. For simplicity, in our tests we choose $L=M=0$. The nature of such perturbations is unspecified: \cite{reggewheeler1957} attribute these effects to distant, massive objects exerting gravitational attraction; \cite{giddingspsaltis2016} explore perturbations induced by quantum effects manifesting on a global scale around the event horizon, by introducing specific forms of the functions $F(r),K(r)$. In both cases, the macroscopic effect is a change in the shape of the shadow of the central object that can be observed e.g. by instruments such as the Event Horizon Telescope (EHT). In \cite{johanssen2011b} a similar approach is considered for axisymmetric and time-independent perturbations to the Kerr metric, in order to explore a quasi-Kerr black hole showing violations of the No-Hair theorem (\citealt{psaltis2012}; \citealt{vincent2016}; \citealt{psaltis2016}) or to describe geodesics around neutron stars (\citealt{baubock2012}).

\begin{figure}[!h]
\centering
\subfloat[Unperturbed Schwarzschild spacetime.]{\includegraphics[scale=.5]{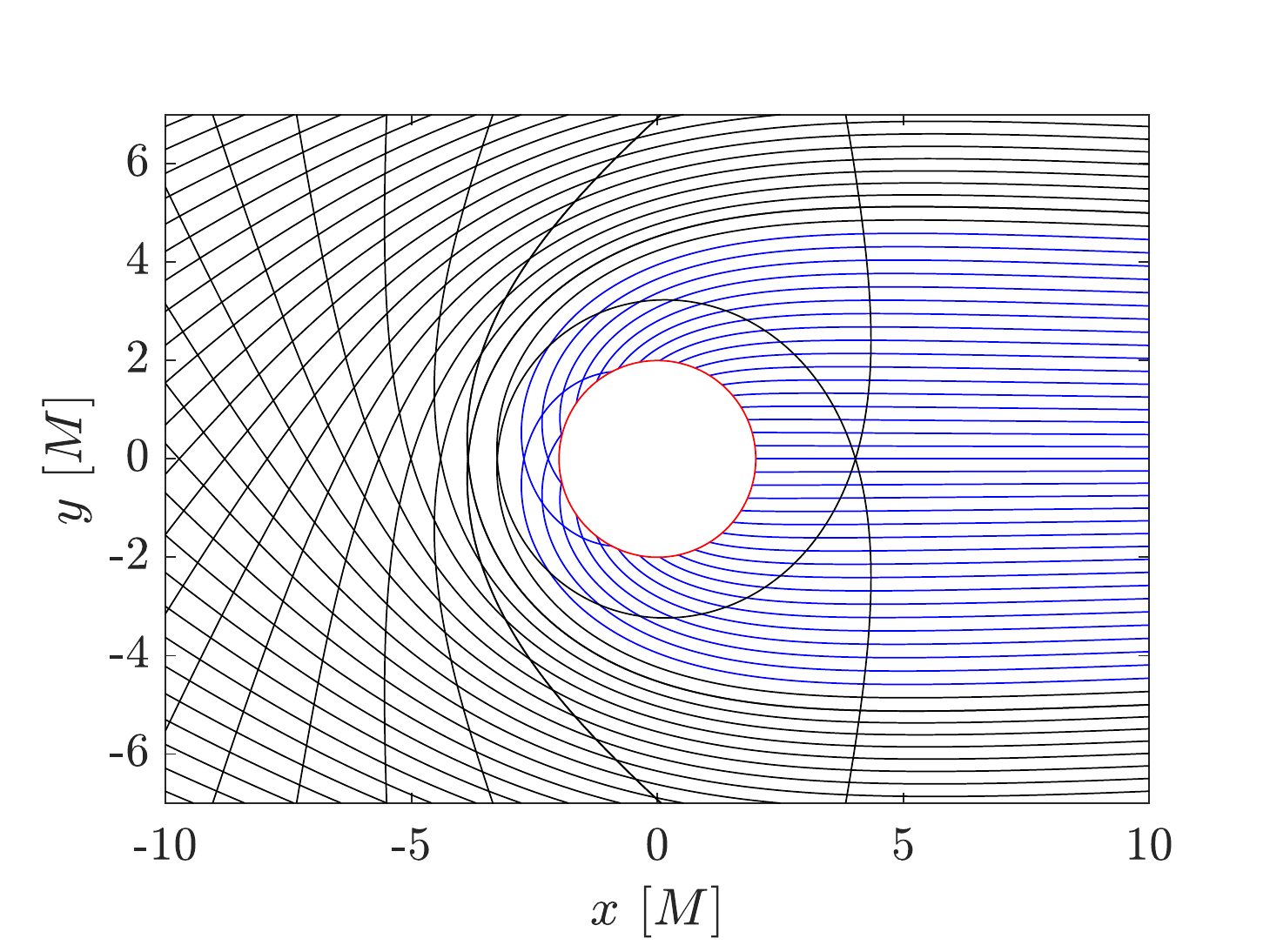}}
\subfloat[Perturbed Schwarzschild spacetime.]{\includegraphics[scale=.5]{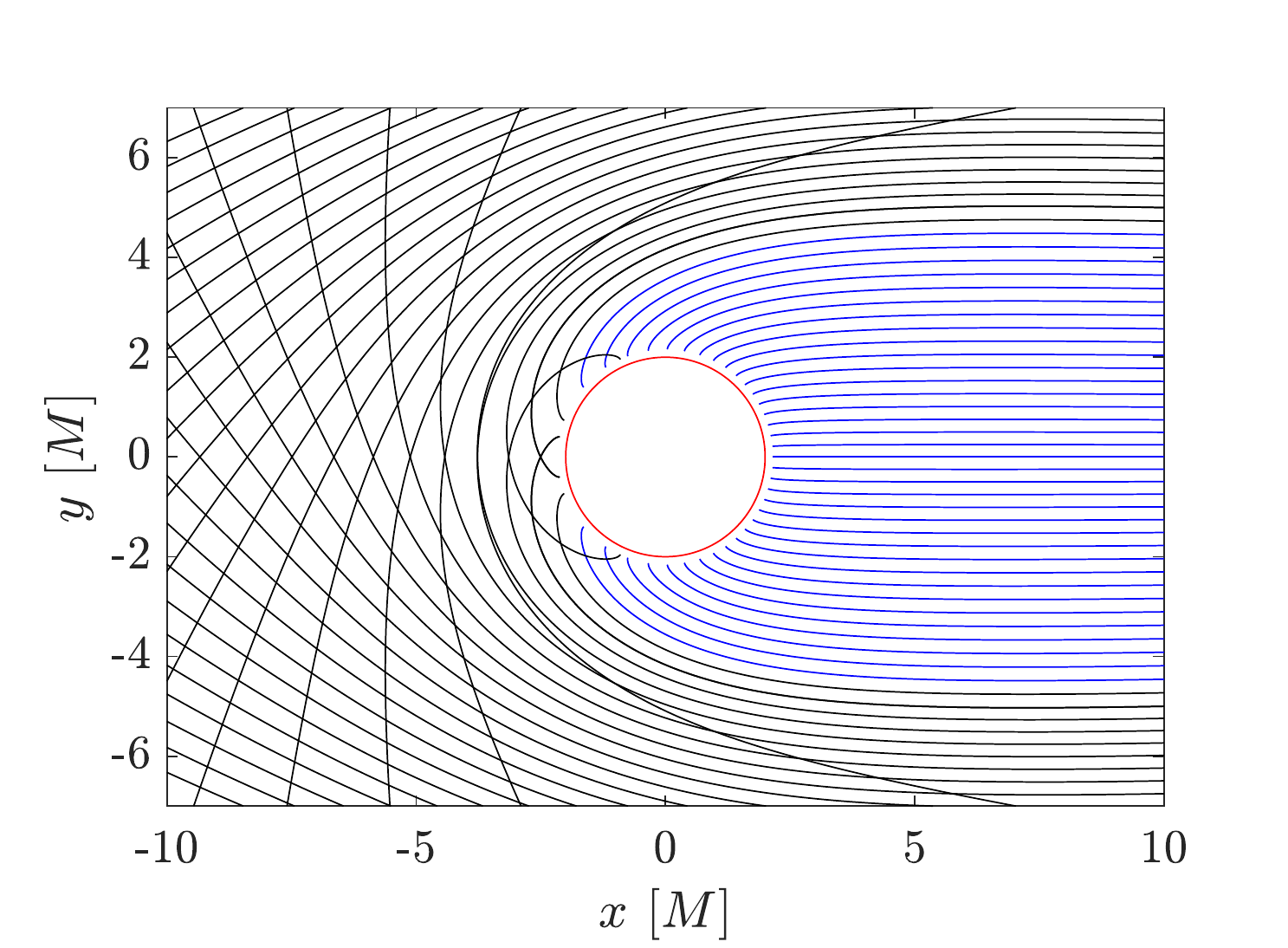}}
\caption{Trajectories of photons in the $r-\varphi$ plane in unperturbed Schwarzschild spacetime (left-hand panel) and in the perturbed metric from equation (\ref{eq:schwarzpert}) (right-hand panel). The light rays are initialized on parallel trajectories at the right boundary and approach the central object, either falling in below the Schwarzschild radius (red circle) or being deflected from the straight path. The blue trajectories mark the photons that end up crossing the event horizon in the unperturbed case; the escaping photon trajectories are colored in black. The chosen perturbation of the metric causes some of the photons that escape in the unperturbed case to being captured instead.}
\label{fig:schwarzpert}
\end{figure}

In general, one can expect that perturbations of this kind affect the metric significantly only around the Schwarzschild radius, and then quickly decay away from it. Thus, light rays passing close to $r=r_S$ are likely to be deflected in a different manner than what observed in the unperturbed case. Figure \ref{fig:schwarzpert} exemplifies the resulting effect on simulated photon trajectories in the $r-\varphi$ plane. Here, a number of light rays coming from infinity enter the simulation box from the right boundary. In the unperturbed Schwarzschild metric (left-hand panel), some trajectories (indicated as blue lines) lead to photons falling into the event horizon at $r=r_S$ (marked by a red circle), while others are deflected and escape to infinity (black lines). In the perturbed metric (right-hand panel), some of the black trajectories that would lead to escape in the previous case end up falling into the black hole instead. As expected, the difference between the two cases is confined to a small region around $r_S$.

In these tests, our choice of implementation proves particularly useful, since perturbations in the form (\ref{eq:schwarzpert}) can be included right away without changing the code structure. Note that we can also retain the usual Schwarschild spherical coordinates without the need to convert to Eddington-Finkelstein variables, or monitor signature changes in the metric like in \cite{giddingspsaltis2016}. Another advantage of our implementation is that there is no restriction on the form of the perturbations, both axisymmetric and non-axisymmetric perturbations can be added to any metric.

\section{Discussion and summary}
\label{sec:discussionsummary}

We presented a versatile algorithm for the numerical integration of geodesics in general relativity, based on the 3+1 ADM formalism. In this framework, we compared the performance of three different numerical integrators, namely a standard fourth-order explicit Runge-Kutta (RK4) scheme, a second-order implicit midpoint rule (IMR) scheme, and a new, second-order implicit Hamiltonian scheme. The new scheme is exactly energy-conserving, since it is based on the preservation of the underlying Hamiltonian to numerical round-off accuracy. We applied all schemes to a number of standard and non-standard spacetimes, simulating both photon and massive particle trajectories.

For geodesics in Schwarzschild and Kerr spacetimes, we observed improvements in energy and position errors when simulating photon trajectories near the event horizon with the new energy-conserving scheme. Regions around the coordinate singularities proved pathological for explicit integrators, since energy errors grow uncontrollably, deviating the simulated light rays from the correct path. Einstein rings around a Schwarzschild black hole and unstable spherical photon orbits around a spinning (Kerr) black hole can be modeled more accurately with the Hamiltonian scheme without reduction of the time step to values that make the computational cost prohibitive. The resulting error is orders of magnitude smaller than that observed for the RK4 and IMR schemes. In the context of ray-tracing and black hole imaging, the new scheme can be applied dynamically while monitoring energy errors for each geodesic calculated with one of the standard schemes (e.g. RK4). In this way, the computational cost of the overall scheme does not increase dramatically, while energy errors are corrected to prevent unphysical outcomes.

As an example, we applied the dynamical scheme outlined above to a ray-tracing calculation in Kerr spacetime. Following the work by \cite{alonso2008}, we initialized $N_{ph}=512^2$ photons in the $\theta=\pi/2$ plane of an extremal Kerr black hole of mass $M=1$, at random positions in the range $r=[r_+,10r_S]$, at randomized directions (we keep track of the seed used in the random number generator for reproducibility). We integrate each photon geodesic and we monitor the escape of the photons from the $r=10r_S$ boundary of the simulated domain. We then calculate the fraction of escaping photons, $N_{esc}/N_{ph}$, obtained with the RK4 method using a fixed time step $\Delta t=5$. Then, we run the same simulation, this time allowing for dynamically switching to the Hamiltonian method in case an energy error larger than a prescribed tolerance is detected. We compare the two outcomes with a reference run that uses the RK4 method with an extremely small time step $\Delta t=0.01$.

The results show that the escape fraction obtained with the RK4 method and $\Delta t=5$ differs from the reference result by $\sim 1\%$. The simulation including the dynamical switch to the Hamiltonian scheme, instead, reproduces the reference result exactly, even though the time step is more than two orders of magnitude larger. To confirm the generality of the results, we run the same analysis by varying the outer boundary of the simulation box from $r=2r_+$ to $r=10r_S$. For the simulations run with RK4 and $\Delta t=5$, we find errors in the computed escape fraction that range from $\sim 0.5\%$ to $\sim 2\%$. The results obtained with the Hamiltonian scheme and the same $\Delta t$, instead, correspond exactly to the reference run in all cases.

Because the switch to the Hamiltonian scheme is treated dynamically within the algorithm, the additional cost is only due to the recalculation of the geodesics that present high energy errors, thus optimizing the global scheme. It must be noted that, even though the results obtained with the RK4 method present only small errors, in production ray-tracing calculations with millions or billions of geodesics such errors may affect a significantly high number of photon paths. In particular, the regions of spacetime that are most affected by energy errors are those surrounding the event horizon, as shown in Section \ref{sec:tests}, which are also the most interesting zones due to the presence of high-order Einstein rings created by frame-dragging. While those regions of spacetime far away from the metric singularities are efficiently treated by standard methods (such as less expensive RK4 schemes), the possibility of studying more complicated features with an acceptable computational effort (e.g. with the Hamiltonian method, as shown) is certainly an advantage.

\begin{figure}[!h]
\centering
\subfloat[Observer's view in flat spacetime.]{\includegraphics[scale=.17]{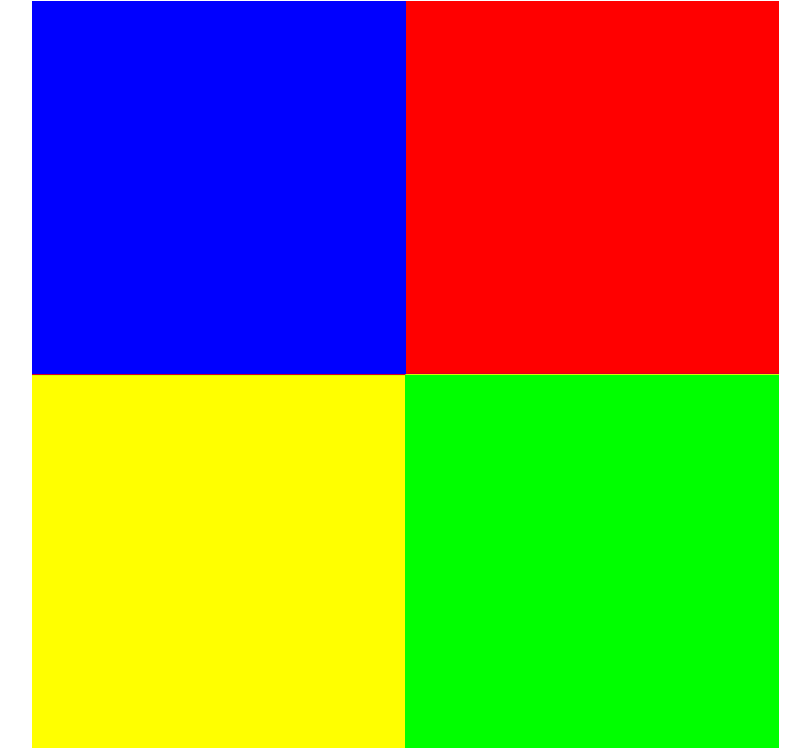}} \quad
\subfloat[Black hole shadow observed from the equatorial plane.]{\includegraphics[scale=.17]{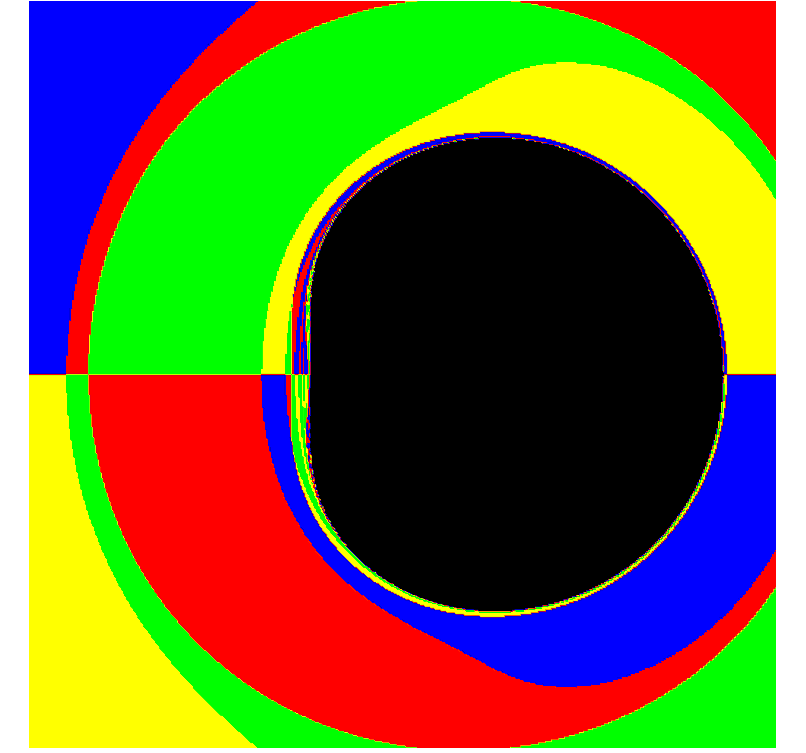}} \quad
\subfloat[Shadow of a black hole surrounded by a white accretion disk, observed at an angle $\pi/18$ above the equatorial plane.]{\includegraphics[scale=.165]{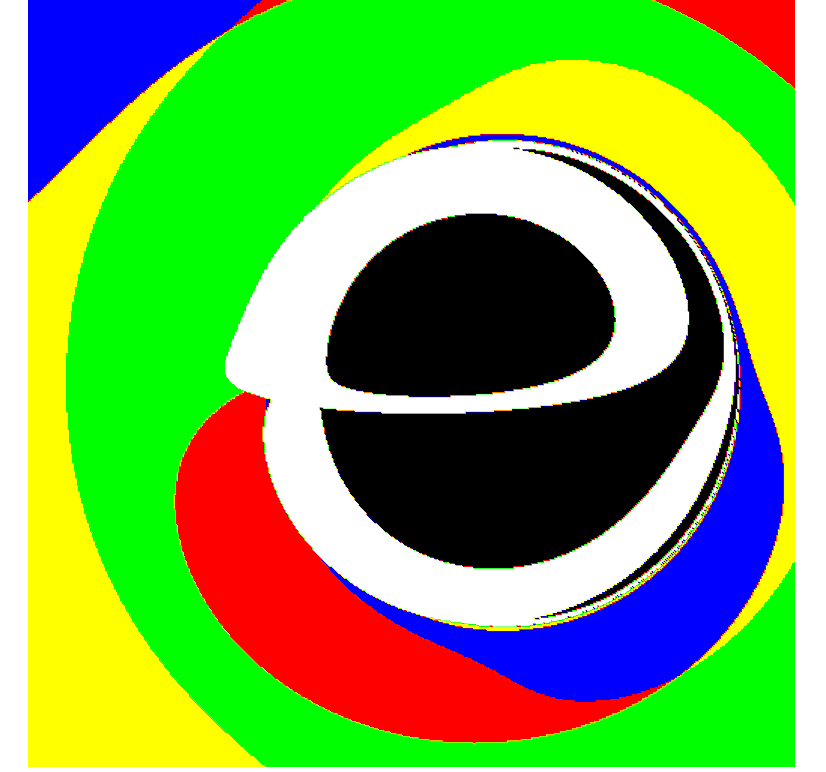}}
\caption{An observer's view of a distant four-color background (left-hand panel) is distorted by the presence of an extremal Kerr black hole (central and right-hand panels), rotating from left to right. The central panel shows the black hole shadow as observed from the $\theta=\pi/2$ equatorial plane. The black hole spin distorts the  shadow such that the observed shape of the event horizon is not symmetric with respect to the axis of rotation. Einstein rings of higher and higher order are visible on one side of the shadow, due to frame dragging. In the right-hand panel, the black hole is surrounded by an infinitely think accretion disk (in white), and observed from an angle $\pi/18$ above the equatorial plane. The region of the accretion disk that would be hidden behind the event horizon is instead visible due to gravitational bending of light rays. This picture was obtained by ray-tracing photon geodesics from the observer's eye (a camera with resolution of $512^2$ pixels) back to the point in space where such photons were emitted.}
\label{fig:shadow}
\end{figure}

High-precision numerical integrators are also intrinsically necessary for the study of those spacetimes that are non-integrable (e.g. the dihole solution from Section 5) and therefore do not present analytic solutions, while being characterized by interesting features such as fractal-like Einstein rings around the black hole shadow (see e.g. \citealt{wang2018}). The astrophysical relevance of such calculations is clear in the context of comparisons with upcoming observations of the black hole shadow, which will shed light on the correctness of present theories of gravity. An example of such a calculation (obtained with the RK4 scheme with $\Delta t=0.01$ is shown in Figure \ref{fig:shadow}, where an extremal Kerr black hole is placed between an observer and a four-color background. The left-hand panel shows the observer's view in flat spacetime, in absence of black holes. In the central panel, the observed image is highly distorted due to the gravitational bending of the light rays. In the right-hand panel, the black hole is surrounded by an infinitely thin accretion disk. The observer is allowed to see parts of the accretion disk that would be normally out of sight, being covered by the event horizon. The process of numerically imaging such objects accurately is of high relevance for comparing 
to observational data and provide support to theoretical findings (see e.g. \citealt{mizuno2018}).

The choice of 3+1 splitting the equations of motion proves particularly useful in designing a generic algorithm for any spacetime. In fact, only the 3+1 metric functions $\alpha$, $\beta^i$, and $\gamma^{ij}$ have to be provided as input. This allows for the simulation of free particle motion in both standard (Schwarzschild and Kerr) and nonstandard spacetimes without the need to restructure the code or customize the algorithm for specific metrics. In order to exemplify the potential of this choice of implementation, we simulated the motion of massless and massive particles around a Morris-Thorne wormhole, a Reissner-Nordstr\"{o}m dihole system, and a perturbed Schwarzschild black hole. In all cases, the only required effort was to provide the metric functions initially.

In the perspective of extending our studies to the simulation of general relativistic charged particles in EM fields, we need to consider the properties of each integration scheme when applied to a more complex physical situation. Numerical integrators for the motion of relativistic charged particles in EM fields in curved spacetime are required for particle tracing in plasma simulations (e.g. GRMHD codes, or future GR Particle-in-Cell kinetic codes), in order to fully capture the general relativistic effects induced on the plasma flows around compact objects. It is well known that the RK4 scheme is non-ideal for specific applications, e.g. gyrating motion in a magnetic field, due to the unbounded accumulation of truncation errors (\citealt{Qin_2013}). It is reasonable to suspect that in the corresponding GR case, the same scheme performs poorly, considering the non-ideal performance we detected, in some cases, for geodesic motion alone. Alternative, explicit schemes exist and have been widely studied and applied to special relativistic charged particles (see \citealt{ripperda2018} for an extensive review on the subject). However, the extension of such schemes to general relativity seems to be non-straightforward and requires further investigation. Despite the lack of exact energy conservation, implicit, symplectic integrators such as the IMR scheme have been successfully applied to special relativistic regimes (\citealt{HC_2017}). This motivates the possibility of choosing such a scheme for future applications to charged particle motion in curved spacetimes. Energy conservation can be achieved with a new Hamiltonian scheme, a feature that makes it especially attractive for the purpose of simulating particles in such complicated physical systems. In fact, the energy-conserving character of the scheme can be retained regardless of the underlying physics, provided that the system can be described with a Hamiltonian. Hence, the inclusion of Lorentz force in addition to the geodesic motion is straightforward, opening the path for obtaining highly accurate simulations of relativistic charged particle dynamics around compact objects.

\section*{Acknowledgements}
BR and FB were supported by projects GOA/2015-014 (2014-2018 KU Leuven) and the Interuniversity Attraction Poles Programme by the Belgian Science Policy Office (IAP P7/08 CHARM). FB is also supported by the Research Fund KU Leuven and Space Weaves RUN project. LS acknowledges support from DoE DE-SC0016542, NASA Fermi NNX-16AR75G, NASA ATP NNX-17AG21G, NSF ACI-1657507, and NSF AST- 1716567. The authors would like to thank Rony Keppens, Gianni Lapenta, Oliver Porth, Anatoly Spitkovsky, Leo Stein, Bert Vercnocke, and Thomas Hertog for useful discussions and CK Chan, Ziri Younsi, and Thomas M\"{u}ller for explaining and sharing their codes and data with us.

%%%%%%%%%%%%%%%%%%%%%%%%%%%%%%%%%%%%%%%%%%%%%%%%%%

%%%%%%%%%%%%%%%%% APPENDICES %%%%%%%%%%%%%%%%%%%%%

\appendix

\section{Discrete energy-conserving Hamiltonian scheme}
\label{app:ham}

Given the Hamiltonian
\begin{equation}
 H(x^i,u_i)=\alpha(\gamma^{jk} u_j u_k+\epsilon)^{1/2} - \beta^j u_j,
 \label{eq:hamiltonianapp}
\end{equation}
for a system of 3 equations for $x^i$ and 3 equations for $u_i$, a suitable discretization-averaging of the Hamiltonian equations reads
\begin{equation}
 \begin{aligned}
 \frac{x^{1,n+1}-x^{1,n}}{\Delta t} = &\frac{1}{6}\left[\frac{H(x^{1,n+1},x^{2,n},x^{3,n},u_1^{n+1},u_2^{n},u_3^{n})-H(x^{1,n+1},x^{2,n},x^{3,n},u_1^{n},u_2^{n},u_3^{n})}{u_1^{n+1}-u_1^n} \right. \\ 
 & +\frac{H(x^{1,n+1},x^{2,n+1},x^{3,n+1},u_1^{n+1},u_2^{n+1},u_3^{n+1})-H(x^{1,n+1},x^{2,n+1},x^{3,n+1},u_1^{n},u_2^{n+1},u_3^{n+1})}{u_1^{n+1}-u_1^n} \\
 & +\frac{H(x^{1,n},x^{2,n+1},x^{3,n+1},u_1^{n+1},u_2^{n+1},u_3^{n+1})-H(x^{1,n},x^{2,n+1},x^{3,n+1},u_1^{n},u_2^{n+1},u_3^{n+1})}{u_1^{n+1}-u_1^n} \\
 & +\frac{H(x^{1,n+1},x^{2,n},x^{3,n+1},u_1^{n+1},u_2^{n},u_3^{n+1})-H(x^{1,n+1},x^{2,n},x^{3,n+1},u_1^{n},u_2^{n},u_3^{n+1})}{u_1^{n+1}-u_1^n} \\
 & +\frac{H(x^{1,n},x^{2,n},x^{3,n},u_1^{n+1},u_2^{n},u_3^{n})-H(x^{1,n},x^{2,n},x^{3,n},u_1^{n},u_2^{n},u_3^{n})}{u_1^{n+1}-u_1^n} \\
 & \left. +\frac{H(x^{1,n},x^{2,n},x^{3,n+1},u_1^{n+1},u_2^{n},u_3^{n+1})-H(x^{1,n},x^{2,n},x^{3,n+1},u_1^{n},u_2^{n},u_3^{n+1})}{u_1^{n+1}-u_1^n}\right],
 \end{aligned}
 \label{eq:poshamdisc1}
\end{equation}
\begin{equation}
 \begin{aligned}
 \frac{x^{2,n+1}-x^{2,n}}{\Delta t} = &\frac{1}{6}\left[\frac{H(x^{1,n+1},x^{2,n+1},x^{3,n},u_1^{n+1},u_2^{n+1},u_3^{n})-H(x^{1,n+1},x^{2,n+1},x^{3,n},u_1^{n+1},u_2^{n},u_3^{n})}{u_2^{n+1}-u_2^n} \right. \\
 & +\frac{H(x^{1,n},x^{2,n+1},x^{3,n},u_1^{n},u_2^{n+1},u_3^{n})-H(x^{1,n},x^{2,n+1},x^{3,n},u_1^{n},u_2^{n},u_3^{n})}{u_2^{n+1}-u_2^n} \\
 & +\frac{H(x^{1,n},x^{2,n},x^{3,n},u_1^{n},u_2^{n+1},u_3^{n})-H(x^{1,n},x^{2,n},x^{3,n},u_1^{n},u_2^{n},u_3^{n})}{u_2^{n+1}-u_2^n} \\
 & +\frac{H(x^{1,n+1},x^{2,n+1},x^{3,n+1},u_1^{n+1},u_2^{n+1},u_3^{n+1})-H(x^{1,n+1},x^{2,n+1},x^{3,n+1},u_1^{n+1},u_2^{n},u_3^{n+1})}{u_2^{n+1}-u_2^n} \\
 & +\frac{H(x^{1,n+1},x^{2,n},x^{3,n},u_1^{n+1},u_2^{n+1},u_3^{n})-H(x^{1,n+1},x^{2,n},x^{3,n},u_1^{n+1},u_2^{n},u_3^{n})}{u_2^{n+1}-u_2^n} \\
 & \left. +\frac{H(x^{1,n+1},x^{2,n},x^{3,n+1},u_1^{n+1},u_2^{n+1},u_3^{n+1})-H(x^{1,n+1},x^{2,n},x^{3,n+1},u_1^{n+1},u_2^{n},u_3^{n+1})}{u_2^{n+1}-u_2^n}\right],
 \end{aligned}
 \label{eq:poshamdisc2}
\end{equation}
\begin{equation}
 \begin{aligned}
 \frac{x^{3,n+1}-x^{3,n}}{\Delta t} = &\frac{1}{6}\left[\frac{H(x^{1,n+1},x^{2,n+1},x^{3,n+1},u_1^{n+1},u_2^{n+1},u_3^{n+1})-H(x^{1,n+1},x^{2,n+1},x^{3,n+1},u_1^{n+1},u_2^{n+1},u_3^{n})}{u_3^{n+1}-u_3^n} \right. \\
 & +\frac{H(x^{1,n},x^{2,n+1},x^{3,n+1},u_1^{n},u_2^{n+1},u_3^{n+1})-H(x^{1,n},x^{2,n+1},x^{3,n+1},u_1^{n},u_2^{n+1},u_3^{n})}{u_3^{n+1}-u_3^n} \\
 & +\frac{H(x^{1,n},x^{2,n+1},x^{3,n},u_1^{n},u_2^{n+1},u_3^{n+1})-H(x^{1,n},x^{2,n+1},x^{3,n},u_1^{n},u_2^{n+1},u_3^{n})}{u_3^{n+1}-u_3^n} \\
 & +\frac{H(x^{1,n},x^{2,n},x^{3,n+1},u_1^{n},u_2^{n},u_3^{n+1})-H(x^{1,n},x^{2,n},x^{3,n+1},u_1^{n},u_2^{n},u_3^{n})}{u_3^{n+1}-u_3^n} \\
 & +\frac{H(x^{1,n+1},x^{2,n+1},x^{3,n},u_1^{n+1},u_2^{n+1},u_3^{n+1})-H(x^{1,n+1},x^{2,n+1},x^{3,n},u_1^{n+1},u_2^{n+1},u_3^{n})}{u_3^{n+1}-u_3^n} \\
 & \left. +\frac{H(x^{1,n},x^{2,n},x^{3,n},u_1^{n},u_2^{n},u_3^{n+1})-H(x^{1,n},x^{2,n},x^{3,n},u_1^{n},u_2^{n},u_3^{n})}{u_3^{n+1}-u_3^n}\right],
 \end{aligned}
 \label{eq:poshamdisc3}
\end{equation}
\begin{equation}
 \begin{aligned}
 \frac{u_1^{n+1}-u_1^n}{\Delta t} = &\frac{1}{6}\left[\frac{H(x^{1,n+1},x^{2,n},x^{3,n},u_1^{n},u_2^{n},u_3^{n})-H(x^{1,n},x^{2,n},x^{3,n},u_1^{n},u_2^{n},u_3^{n})}{x^{1,n+1}-x^{1,n}} \right. \\
 & +\frac{H(x^{1,n+1},x^{2,n+1},x^{3,n+1},u_1^{n},u_2^{n+1},u_3^{n+1})-H(x^{1,n},x^{2,n+1},x^{3,n+1},u_1^{n},u_2^{n+1},u_3^{n+1})}{x^{1,n+1}-x^{1,n}} \\
 & +\frac{H(x^{1,n+1},x^{2,n+1},x^{3,n+1},u_1^{n+1},u_2^{n+1},u_3^{n+1})-H(x^{1,n},x^{2,n+1},x^{3,n+1},u_1^{n+1},u_2^{n+1},u_3^{n+1})}{x^{1,n+1}-x^{1,n}} \\
 & +\frac{H(x^{1,n+1},x^{2,n},x^{3,n+1},u_1^{n},u_2^{n},u_3^{n+1})-H(x^{1,n},x^{2,n},x^{3,n+1},u_1^{n},u_2^{n},u_3^{n+1})}{x^{1,n+1}-x^{1,n}} \\
 & +\frac{H(x^{1,n+1},x^{2,n},x^{3,n},u_1^{n+1},u_2^{n},u_3^{n})-H(x^{1,n},x^{2,n},x^{3,n},u_1^{n+1},u_2^{n},u_3^{n})}{x^{1,n+1}-x^{1,n}} \\
 & \left. +\frac{H(x^{1,n+1},x^{2,n},x^{3,n+1},u_1^{n+1},u_2^{n},u_3^{n+1})-H(x^{1,n},x^{2,n},x^{3,n+1},u_1^{n+1},u_2^{n},u_3^{n+1})}{x^{1,n+1}-x^{1,n}}\right],
 \end{aligned}
 \label{eq:velhamdisc1}
\end{equation}
\begin{equation}
 \begin{aligned}
 \frac{u_2^{n+1}-u_2^n}{\Delta t} = &\frac{1}{6}\left[\frac{H(x^{1,n+1},x^{2,n+1},x^{3,n},u_1^{n+1},u_2^{n},u_3^{n})-H(x^{1,n+1},x^{2,n},x^{3,n},u_1^{n+1},u_2^{n},u_3^{n})}{x^{2,n+1}-x^{2,n}} \right. \\ 
 & +\frac{H(x^{1,n},x^{2,n+1},x^{3,n},u_1^{n},u_2^{n},u_3^{n})-H(x^{1,n},x^{2,n},x^{3,n},u_1^{n},u_2^{n},u_3^{n})}{x^{2,n+1}-x^{2,n}} \\
 & +\frac{H(x^{1,n},x^{2,n+1},x^{3,n},u_1^{n},u_2^{n+1},u_3^{n})-H(x^{1,n},x^{2,n},x^{3,n},u_1^{n},u_2^{n+1},u_3^{n})}{x^{2,n+1}-x^{2,n}} \\
 & +\frac{H(x^{1,n+1},x^{2,n+1},x^{3,n+1},u_1^{n+1},u_2^{n},u_3^{n+1})-H(x^{1,n+1},x^{2,n},x^{3,n+1},u_1^{n+1},u_2^{n},u_3^{n+1})}{x^{2,n+1}-x^{2,n}} \\
 & +\frac{H(x^{1,n+1},x^{2,n+1},x^{3,n},u_1^{n+1},u_2^{n+1},u_3^{n})-H(x^{1,n+1},x^{2,n},x^{3,n},u_1^{n+1},u_2^{n+1},u_3^{n})}{x^{2,n+1}-x^{2,n}} \\
 & \left. +\frac{H(x^{1,n+1},x^{2,n+1},x^{3,n+1},u_1^{n+1},u_2^{n+1},u_3^{n+1})-H(x^{1,n+1},x^{2,n},x^{3,n+1},u_1^{n+1},u_2^{n+1},u_3^{n+1})}{x^{2,n+1}-x^{2,n}}\right],
 \end{aligned}
 \label{eq:velhamdisc2}
\end{equation}
\begin{equation}
 \begin{aligned}
 \frac{u_3^{n+1}-u_3^n}{\Delta t} = &\frac{1}{6}\left[\frac{H(x^{1,n+1},x^{2,n+1},x^{3,n+1},u_1^{n+1},u_2^{n+1},u_3^{n})-H(x^{1,n+1},x^{2,n+1},x^{3,n},u_1^{n+1},u_2^{n+1},u_3^{n})}{x^{3,n+1}-x^{3,n}} \right. \\
 & +\frac{H(x^{1,n},x^{2,n+1},x^{3,n+1},u_1^{n},u_2^{n+1},u_3^{n})-H(x^{1,n},x^{2,n+1},x^{3,n},u_1^{n},u_2^{n+1},u_3^{n})}{x^{3,n+1}-x^{3,n}} \\
 & +\frac{H(x^{1,n},x^{2,n+1},x^{3,n+1},u_1^{n},u_2^{n+1},u_3^{n+1})-H(x^{1,n},x^{2,n+1},x^{3,n},u_1^{n},u_2^{n+1},u_3^{n+1})}{x^{3,n+1}-x^{3,n}} \\
 & +\frac{H(x^{1,n},x^{2,n},x^{3,n+1},u_1^{n},u_2^{n},u_3^{n})-H(x^{1,n},x^{2,n},x^{3,n},u_1^{n},u_2^{n},u_3^{n})}{x^{3,n+1}-x^{3,n}} \\
 & +\frac{H(x^{1,n+1},x^{2,n+1},x^{3,n+1},u_1^{n+1},u_2^{n+1},u_3^{n+1})-H(x^{1,n+1},x^{2,n+1},x^{3,n},u_1^{n+1},u_2^{n+1},u_3^{n+1})}{x^{3,n+1}-x^{3,n}} \\
 & \left. +\frac{H(x^{1,n},x^{2,n},x^{3,n+1},u_1^{n},u_2^{n},u_3^{n+1})-H(x^{1,n},x^{2,n},x^{3,n},u_1^{n},u_2^{n},u_3^{n+1})}{x^{3,n+1}-x^{3,n}}\right].
 \end{aligned}
 \label{eq:velhamdisc3}
\end{equation}

One peculiarity of the equations above is the presence of the position and momentum increments, $x^{i,n+1}-x^{i,n}$ and $u_i^{n+1}-u_i^n$ respectively, in the denominator of the right-hand side. It is clear that difficulties in the solution can arise whenever such increments tend to zero. Care must be taken in handling this issue throughout the computation, as the results can get severely affected by the behavior of the solution around these critical points. In the worst case, the iterative solution may fail to converge, ruining the computation.

In order to avoid this kind of numerical singularities in the system of equations (\ref{eq:poshamdisc1}-\ref{eq:velhamdisc3}), we can rewrite the difference equations in a more convenient form. Without loss of generality, let us consider the discrete equation (\ref{eq:poshamdisc1}). The right-hand side is composed of six difference terms, each one being the finite increment of the Hamiltonian between $u_1^{n+1}$ and $u_1^n$, with the other variables ($x^1$, $x^2$, $x^3$, $u_2$, $u_3$) evaluated at specific combinations of the two time levels. Such a combination is fixed between the two elements of each difference term. Hence, each difference term actually evaluates the ratio
\begin{equation}
 \frac{H(u_1^{n+1},...)-H(u_1^n,...)}{u^{n+1}_1-u^n_1},
 \label{eq:hamincpos}
\end{equation}
where $...$ indicates a combination of the other variables, representing a suitable averaging for the Hamiltonian-preserving scheme. Note that for each difference term the computation effectively reduces to the evaluation of an incremental ratio of the form 
\begin{equation}
 \frac{f(x+h)-f(x)}{h},
\end{equation}
for a generic function $f(x)$ and an arbitrary increment $h$. Each of the six difference terms can be expanded by substituting the definition of the Hamiltonian (\ref{eq:hamiltonianapp}). After a series of manipulations, the general incremental ratio (\ref{eq:hamincpos}) can be written as
\begin{equation}
 \frac{H(u_1^{n+1},...)-H(u_1^n,...)}{u^{n+1}_1-u^n_1} = \alpha\frac{\gamma^{11}(u_1^{n+1}+u_1^n)
 +2\gamma^{12}u_2+2\gamma^{13}u_3}{\sqrt{U_1^{n+1}+\epsilon}+\sqrt{U_1^{n}+\epsilon}} -\beta^1,
\end{equation}
where
\begin{equation}
 U^{n+1}_1=\gamma^{11}(u_1^{n+1})^2+2\gamma^{12}u_1^{n+1}u_2+2\gamma^{13}u_1^{n+1}u_3+\gamma^{22}(u_2)^2+\gamma^{33}(u_3)^2+\gamma^{23}u_2 u_3,
\end{equation}
\begin{equation}
 U^{n}_1=\gamma^{11}(u_1^{n})^2+2\gamma^{12}u_1^{n}u_2+2\gamma^{13}u_1^{n}u_3+\gamma^{22}(u_2)^2+\gamma^{33}(u_3)^2+\gamma^{23}u_2 u_3.
\end{equation}
Note that the increment in $u_1$ vanishes, in the process, from the denominator of equation \ref{eq:poshamdisc1}, hence the singularity disappears with this form of the equation. The process can be repeated for each difference term and for equations (\ref{eq:poshamdisc2}) and (\ref{eq:poshamdisc3}). The generalization of the above procedure yields
\begin{equation}
 \frac{x^{i,n+1}-x^{i,n}}{\Delta t} = \sum^6\alpha\frac{\gamma^{ii}(u_i^{n+1}+u_i^n)
 +2\gamma^{ij}u_j+2\gamma^{ik}u_k}{\sqrt{U_i^{n+1}+\epsilon}+\sqrt{U_i^{n}+\epsilon}} -\beta^i.
 \label{eq:poshamdiscsimple}
\end{equation}
This form of the difference equations (\ref{eq:poshamdisc1}-\ref{eq:poshamdisc3}) avoids singularities in $u_i^{n+1}-u_i^n$, and is therefore more convenient to use in an iterative solution procedure.

A similar simplification procedure applied to equations (\ref{eq:velhamdisc1}-\ref{eq:velhamdisc3}) yields
\begin{equation}
\begin{aligned}
 \frac{u^{n+1}_i-u^{n}_i}{\Delta t} = & \sum^6 -\frac{1}{2}\left(\sqrt{\gamma^{lm}(x^{i,n+1},...)u_l u_m+\epsilon}+\sqrt{\gamma^{lm}(x^{i,n},...)u_l u_m+\epsilon}\right) \frac{\alpha(x^{i,n+1},...)-\alpha(x^{i,n},...)}{x^{i,n+1}-x^{i,n}} \\
  & -\frac{1}{2}\frac{\alpha(x^{i,n+1},...)+\alpha(x^{i,n},...)}{\sqrt{\gamma^{lm}(x^{i,n+1},...)u_l u_m+\epsilon}+\sqrt{\gamma^{lm}(x^{i,n},...)u_l u_m+\epsilon}}u_l u_m \frac{\gamma^{lm}(x^{i,n+1},...)-\gamma^{lm}(x^{i,n},...)}{x^{i,n+1}-x^{i,n}} \\
 & + u_l\frac{\beta^l(x^{i,n+1},...)-\beta^l(x^{i,n},...)}{x^{i,n+1}-x^{i,n}}.
  \label{eq:velhamdiscsimple}
\end{aligned}
\end{equation}
Here, the factor $1/(x^{i,n+1}-x^{i,n})$ does not vanish, thus problems related to singularities may still arise. However, by looking carefully at the quotients in the equation above, it is clear that, for a sufficiently small difference $x^{i,n+1}-x^{i,n}$, this expression reduces to
\begin{equation}
\begin{aligned}
 \frac{u^{n+1}_i-u^{n}_i}{\Delta t} = & \sum^6 -\frac{1}{2}\left(\sqrt{\gamma^{lm}(x^{i,n+1},...)u_l u_m+\epsilon}+\sqrt{\gamma^{lm}(x^{i,n},...)u_l u_m+\epsilon}\right) \partial_i\alpha(x^{i,n},...) \\
  & -\frac{1}{2}\frac{\alpha(x^{i,n+1},...)+\alpha(x^{i,n},...)}{\sqrt{\gamma^{lm}(x^{i,n+1},...)u_l u_m+\epsilon}+\sqrt{\gamma^{lm}(x^{i,n},...)u_l u_m+\epsilon}}u_l u_m \partial_i\gamma^{lm}(x^{i,n},...) \\
 & + u_l\partial_i\beta^l(x^{i,n},...).
  \label{eq:velhamcontsimple}
\end{aligned}
\end{equation}
hence the solution procedure must be handled by substituting (\ref{eq:velhamdiscsimple}) with (\ref{eq:velhamcontsimple}) when the difference $x^{i,n+1}-x^{i,n}<\delta$, where $\delta$ is a prescribed tolerance. A typical choice for such a threshold is $\delta\sim\sqrt{\varepsilon}$, where $\varepsilon$ is the chosen round-off precision (\citealt{press}).

\section{Computational cost of the numerical integration schemes}
\label{app:compcost}
Here, we briefly discuss the computational cost of the schemes presented in Section \ref{sec:schemes}. Particularly, we are interested in assessing the additional complexity of implicit schemes (IMR and Hamiltonian) as compared to the usual cost of a standard RK4 implementation. We do not focus on optimized performance, hence we make no reference to specialized architectures and/or parallel implementations for simulations of particle ensembles. As a reliable measure of comparison, we can think of counting the number of times the right-hand side of our system of equations (\ref{eq:geodesic3p1x}-\ref{eq:geodesic3p1u}) is evaluated numerically. This is an operation that needs to be done in all methods, hence it provides a good reference cost to evaluate the performance of each scheme.

As outlined in Section \ref{sec:rk4}, a standard RK4 algorithm requires the evaluation of the right-hand side of the model ODE (\ref{eq:discode}) four times per time step. In our case, we have 3 equations for the 3-position $x^i$ and 3 equations for the 3-momentum $u_i$. The total cost is thus 24 evaluations/time step.

The standard IMR scheme requires the iterative solution of the nonlinear equation (\ref{eq:IMR}). In our case, this corresponds to a system of 6 equations to be evaluated, at each time step, a number of times corresponding to the iterations required to reach the chosen tolerance. Additionally, the solution procedure involves a matrix-vector multiplication between the inverse Jacobian of the nonlinear system and the vector of residual functions. Assuming that the Jacobian can be pre-computed, and that each of its elements has the same complexity of evaluation of the right-hand side of each nonlinear equation, the total cost is $(6^2+6)\times n_N$, where $n_N$ is the number of Newton iterations. In our tests, we observe that 3-4 iterations are usually sufficient to reach a prescribed absolute tolerance of $10^{-14}$. If we take the conservative value $n_N=5$, we obtain a total cost of 210 evaluations/time step, roughly 10 times more than the RK4 scheme.

The Hamiltonian method presented above evaluates, at each time step and for each nonlinear iteration, 6 difference terms on the right-hand side of each nonlinear equation. As outlined in Appendix \ref{app:ham}, each such term is roughly twice as complex as a single right-hand side of equations (\ref{eq:geodesic3p1x}-\ref{eq:geodesic3p1u}) (it contains approximately twice as many terms). As a consequence, the Jacobian of such a system requires 6 times as many evaluations for each of its elements, hence bringing the complexity of the solution procedure to $(2\times 6^3+2\times6^2)\times n_N$. Considering again $n_N=5$, we obtain a total cost of 2520 evaluations/time step, roughly 100 times more than the RK4 scheme. However, as explained below, in our experiments we find that the actual cost is much smaller.

Given these estimates, we can compare with the actual runtime from a few reference cases. We refer to the setup for orbit B from Section \ref{sec:kerrph} and run the test case with all three methods up to $t=100$, with $\Delta t=1$. Note that the chosen setup is quite demanding in terms of the necessary calculations of the nonlinear functions and Jacobian to be handled during the Newton iteration, since it involves non-diagonal matrices. In our sample implementation in MATLAB 2017, the measured execution time is 0.086705 s for the RK4 scheme, 0.756144 s for the IMR scheme, and 2.894333 s for the Hamiltonian scheme. While the prediction for the IMR case is fairly accurate (approximately 10 times more expensive than the RK4 method), the Hamiltonian scheme performs much better than foreseen, with an execution time roughly 40 times larger than that measured for the RK4 scheme. The cost estimated above, which is 2.5 times larger than what is actually measured, can be taken as a worst case scenario. Still, our simulations suggest that the actual complexity of calculation is fairly moderate and comparable, as an order of magnitude, to that of the IMR scheme. Finally, we note that the above performance corresponds to the solution of the nonlinear equation with a full Newton scheme for both the IMR and Hamiltonian schemes, which involves the costly evaluation of the associated Jacobian at each time step. If we apply a Picard fixed-point iteration, instead, we measure 0.161458 s and 0.879801 s for the IMR and Hamiltonian schemes, respectively. This corresponds to approximately 2 and 10 times the cost of the RK4 scheme. For the Hamiltonian scheme, this is a very acceptable computational effort given the higher accuracy of the results. More efficient implementations in optimized languages can obviously further reduce the overall cost; additionally, refinements of the RK4 methods (e.g. adaptive time-stepping) involve higher costs, therefore lowering the cost ratio for the Hamiltonian scheme even further. Finally, it should be considered that in typical production runs, a combined approach with dynamic switching between integrators should be adopted, as outlined in Section \ref{sec:discussionsummary}. In this context, non-pathological regions of the spacetime can be treated with less expensive methods, such as the RK4 scheme, while upon detecting large energy errors one could impose the recalculation of the orbit with the Hamiltonian scheme. This approach optimizes the computational cost and the use of the different integration schemes.



\bibliographystyle{apalike}

%\begin{thebibliography}{99}


%\end{thebibliography}

%%%%%%%%%%%%%%%%%%%%%%%%%%%%%%%%%%%%%%%%%%%%%%%%%%

%%%%%%%%%%%%%%%%%%%%%%%%%%%%%%%%%%%%%%%%%%%%%%%%%%

% Don't change these lines
% typesetting comment
\label{lastpage}
\end{document}